\documentclass[acmtog]{acmart}
\acmSubmissionID{763}
\usepackage{subcaption}
\usepackage{booktabs} 

\usepackage{soul}

\usepackage{afterpage}
\usepackage{stfloats}
\usepackage[ruled,linesnumbered]{algorithm2e} 

\SetAlFnt{\small}
\SetAlCapFnt{\small}
\SetAlCapNameFnt{\small}
\SetAlCapHSkip{0pt}

\acmJournal{TOG}
\acmYear{2023} \acmVolume{42} \acmNumber{4} \acmArticle{} \acmMonth{8} \acmPrice{15.00}\acmDOI{10.1145/3592447}

\setcopyright{acmlicensed}

\begin{document}
\title{Composite Motion Learning with Task Control}

\author{Pei Xu}
\orcid{0000-0001-7851-3971}
\affiliation{%
 \institution{Clemson University}
 \streetaddress{1240 Supply Street}
 \city{North Charleston}
 \state{SC}
 \postcode{29405}
 \country{USA}}
\affiliation{
 \institution{Roblox}
 \country{USA}}
\email{peix@clemson.edu}

\author{Xiumin Shang}
\orcid{0009-0003-4833-0674}
\affiliation{%
 \institution{University of California, Merced}
 \country{USA}}
\email{xshang@ucmerced.edu}

\author{Victor Zordan}
\orcid{0000-0002-7309-7013}
\affiliation{
 \institution{Roblox}
 \country{USA}}
\affiliation{%
 \institution{Clemson University}
 \country{USA}}
\email{vbz@clemson.edu}

\author{Ioannis Karamouzas}
\orcid{0009-0000-4315-6556}
\affiliation{%
 \institution{Clemson University}
 \country{USA}}
\email{ioannis@clemson.edu}

\renewcommand\shortauthors{Xu, P. et al}

\begin{abstract}
We present a deep learning method for composite and task-driven motion control for physically simulated characters. In contrast to existing data-driven approaches using reinforcement learning that imitate full-body motions, we learn decoupled motions for specific body parts from multiple reference motions simultaneously and directly by leveraging the use of multiple discriminators in a GAN-like setup. In this process, there is no need of any manual work to produce composite reference motions for learning. Instead, the control policy explores by itself how the composite motions can be combined automatically. We further account for multiple task-specific rewards and train a single, multi-objective control policy. To this end, we propose a novel framework for multi-objective learning that adaptively balances the learning of disparate motions from multiple sources and multiple goal-directed control objectives. In addition, as composite motions are typically augmentations of simpler behaviors, we introduce a sample-efficient method for training composite control policies in an incremental manner, where we reuse a pre-trained policy as the meta policy and train a cooperative policy that adapts the meta one for new composite tasks. We show the applicability of our approach on a variety of challenging multi-objective tasks involving both composite motion imitation and multiple goal-directed control.
\end{abstract}

%
%

\ccsdesc[500]{Computing methodologies~Animation}
\ccsdesc[300]{Computing methodologies~Physical simulation}
\ccsdesc[300]{Computing methodologies~Reinforcement learning}

\keywords{character animation, physics-based control, motion synthesis, reinforcement learning, multi-objective learning, incremental learning, GAN}

%
%

\begin{teaserfigure}
\centering
    \includegraphics[width=\linewidth]{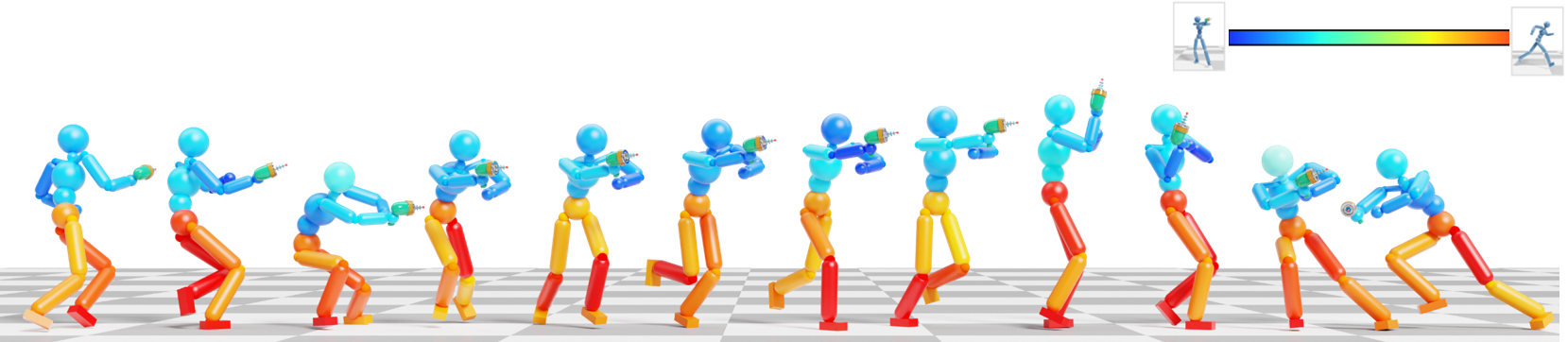}
    \caption{  
    Example of a physically simulated character performing 
    composite motion with locomotion and aiming a weapon. 
    The colors show the automatic mixing of the combined inputs that change dynamically over time based on the state. 
    As indicated in the inset, red denotes body parts that are vital for locomotion while blue for aiming respectively.  
    Our multi-objective approach learns this mixture along with imitation from two disparate reference motions and two goal-directed task rewards for each action.
    }
 \label{fig:teaser}
\end{teaserfigure}

\maketitle

\section{Introduction}
Despite significant advancements in physics-based character control, the majority of existing techniques 
rely on reference data consisting of motion capture recordings of an expert performing the behavior of interest~\cite{peng2018deepmimic,ScaDiver,park2019learning,peng2021amp,iccgan2021,nuttapong2018,peng2022ase,bergamin2019drecon,lee2019scalable}. 
While such reference data is paramount to train motor control policies that lead to natural and robust control, 
in this paper, we are interested in synthesizing  
composite behaviors for physically simulated humanoids by combining \emph{multiple} motion capture reference clips into the training of a single policy. 
Further, we augment these imitation controllers 
with task-specific rewards to train the policy to accomplish specific functional tasks at the same time. 
To this end, we propose a novel multi-objective learning framework that builds \emph{composite} motion behaviors through multiple discriminators, each with its own distinct reference motion as well as task-level control.  Our framework is based on deep reinforcement learning, and allows us to 
adaptively balance the learning of disparate motions from multiple sources and also multiple goal-directed control objectives.

The motivation for this technique is twofold.  First, humans are capable of sophisticated behaviors, including performing multiple tasks simultaneously, such as walking and gesturing or using a mobile phone.  To accomplish this with virtual characters, existing control approaches need to be extended to accommodate the ability to train with multiple objectives as a goal.  Second, with limited exception, most current control frameworks rely on imitation with the style of a behavior being derived from reference motion examples.  Our aim is to be able to combine examples automatically through what we call ``composite motion control'' to avoid the need to continuously seek new example motions for every new permutation of combined behaviors. 
We also explore the ability to add multiple task objectives to support our aim of multi-objective control. 

The core difference of our approach from  existing imitation learning approaches is  decoupling full-body control during training, turning imitation and goal-directed full-body training into a multi-objective learning framework.  To this end, we propose a modification to generative adversarial networks (GANs) to accommodate multiple discriminators (for each subtask in the desired end behavior) and to incorporate the mixing of the  behaviors as a part of the training.  In this way, we sidestep the need to dictate weights for combining the subtasks as well as the need to shape careful reward functions manually for each new composite behavior. 
In addition, as we expect composite motions to often be augmentations from simpler behaviors, we introduce a method for learning composite motion control policies from existing policies through \emph{incremental learning}.  To this end, we train a meta policy, for example for walking, and then train a new policy to \emph{cooperate} with the meta policy, producing a composite motion control policy significantly faster than learning from scratch.  Thus, we can quickly add on to walking new activities from reference data such as punching or waiving, even if we do not have examples of these activities being combined previously with the meta policy.

One naive approach to produce the composite motions we target is to blend motion capture clips to produce a single new motion, and  
perform traditional imitation learning from there. 
This suggested technique may be plausible for simple composite behaviors, like waiving an arm while walking as the two behaviors do not use the same joints, nor do they influence each other greatly, and therefore the blending can be done by simple splicing in a way that is fixed over time. 
Even so,  there is no guarantee of physical plausibility without subsequent training -- and the approach does not scale for more complex behaviors which may have more complicated tradeoffs between body parts used, especially over time.
In contrast, our approach offloads the need to create this weighting as it is produced automatically by the policy as a part of the dictated action. 
Likewise, the output of our system is automatically  guaranteed to be physically valid.  Finally, our approach also has the capability to add task-directed goals, such as walk to a specified location, which is not possible without significant manual effort being added to the naive approach described.

Overall, this paper makes the following contributions:
\vspace{-0.1cm}
\begin{itemize}
    \item We introduce a novel approach for physics-based character control that decouples full-body control in order to learn imitation and task goals from disparate sources and across distinct body parts.
    \item To this end, we extend GAN-style reinforcement learning and introduce a multi-objective learning framework to support multiple discriminators and automatic weighting of imitation and goal-driven subtask rewards.
    \item We propose an incremental learning scheme that uses a meta-policy from an existing behavior to augment the behavior with new subtasks, producing a composite motion control policy that can be learned significantly faster than learning from scratch. Our scheme automatically learns weights across the body that are state dependent in order to effectively mix the original behavior with a new subtask in a temporally dynamic fashion. 
\end{itemize}

\section{Background and Related Work}

\subsection{Physics-Based Character Control}
Developing controllers for physically simulated humanoids has wide applications in computer graphics, robotics, and biomechanics. 
Over the years, a number of trajectory optimization approaches for physics-based control have been proposed that leverage heuristics or feedback rules~\cite{coros2010generalized,wampler2014generalizing,de2009prioritized,ye2010optimal,zordan2014control}, including  open-loop control schemes\cite{liu2010sampling,mordatch2012discovery,liu2015improving}, close-loop feedback control~\cite{mordatch2014combining,da2017tunable} and model predictive control approached~\cite{tassa2012synthesis,tassa2014control,hamalainen2015online,kwon2010control}. 
Given the difficulty in controller design, which often involves multiple optimization objectives, data-driven methods using demonstrations from real humans has also drawn a lot of attention~\cite{da2008simulation,zordan2002motion,kwon2017momentum,lee2010data,yin2007simbicon,liu2012terrain,liu2016guided,muico2009contact,sok2007simulating}. 

In recent years, with the advancement of machine learning techniques, deep reinforcement learning frameworks have gained a lot of popularity 
for training physics-based character controllers.   
While some works~\cite{karpathy2012curriculum,yu2018learning, won2018aerobatics,xie2020allsteps} purely rely on reward functions designed heuristically or using curriculum learning to perform control and encourage the character to act in an expected, human-preferred style,
most recent works leverage motion capture data to perform imitation learning in order to generate high-fidelity, life-like motions. 
DeepLoco~\cite{peng2017deeploco} employs a hierarchical controller to perform walking-style imitation in navigation tasks for a physically simulated character.
DeepMimic~\cite{peng2018deepmimic} combines imitation learning with goal-conditioned learning, and enables a physics-based character to learn a motor skill from a reference motion collected by motion capture or handcrafted by artists.
\citet{nuttapong2018} explore the training of recovery policies that would prevent the character from deviating significantly from the reference motion. 
While the aforementioned works 
rely on a phase variable to synchronize with the reference motion, 
DReCon~\cite{bergamin2019drecon} utilizes a motion matching technique to find the target pose from a collection of reference motions dynamically in response to 
user control input.

Besides relying on direct tracking of reference motions, researchers have offered a number of ways to extend 
the use of reference data in various ways.  For example, 
\citet{park2019learning} leverage the kinematic characteristics of unorganized motions to generate target poses for the control policy to imitate.
UniCon~\cite{wang2020unicon} adopts a similar strategy, where a high-level motion scheduler is employed to provide the target pose for the low-level character controller.
MotionVAE~\cite{ling2020character} employs data-driven generative models using variational autoencoders to generate target motion poses for a reinforcement learning based controller.
A similar model is employed by~\citet{won2022physics} and tested with various goal-directed downstream tasks. 
To ensure synthesis of desired motions, 
these approaches rely on  carefully designed reward functions to assess the controlled character motion.
Drawn from GAIL~\cite{ho2016generative,merel2017learning},   AMP~\cite{peng2021amp} and ICCGAN~\cite{iccgan2021} avoid manually designing reward functions by exploiting the idea of generative adversarial network~(GAN) and relying on a discriminator to obtain the imitation reward for training. 

Beyond the simple use of full-body motions, many works explore motion generation by combining together multiple basic motions with respect to different body parts~\cite{yazaki2015automatic,soga2016body,jang2022motion,jang2008enriching,starke2021neural,alvarado2022generating,liu2018learning}.
However, these works focus on the editing and synthesis of motion animation or using inverse kinematic solvers, and do not work well with current frameworks for controlling physically simulated characters using reinforcement learning.
To date,
existing works for physics-based character control solely focus on the learning of full-body motions. 
As complementary to such works,
in this paper, we target composite motion learning from multiple references without needing to generate any target full-body motion for tasks involving both goal-directed control and imitation control.

\subsection{Training Efficiency}
Characters employed during physics-based control typically are highly articulated with many degrees of freedom defined in continuous action spaces. 
Given the vast feasible choices of action,
controlling so many degrees of freedom is essentially ambiguous, resulting in control problems that are under specified and highly dimensional.
A qualified control policy usually needs millions of samples for training.
The time consumption depends on the exploited algorithms and the motion complexity,
varying from tens of hours to several days.
While some works such as ~\cite{yang2021efficient} explore approaches to speed up the training by improving the reinforcement learning algorithm itself, 
a lot of attention has been recently drawn on 
sample-efficient training by reusing pre-trained policies or action models for fast new motion learning. 
For example, 
many recent approaches employ 
mixture of experts (MoE) models~\cite{peng2019mcp,won2021control,ScaDiver}, where a batch of pre-trained expert policies are exploited to provide primitive actions that are combined by a newly trained policy to generate the final actions.  
Other approaches 
explore using pre-trained latent space models such as variational autoencoders~\cite{ling2020character,won2022physics} and GAN-based models~\cite{peng2022ase} 
to facilitate the training of a control policy.  
In such approaches, the latent space model encapsulates a variety of reference motions and is used by the control policy 
to generate motions for a specific task. 
The works in~\cite{merel2018neural,merel2020catch} 
combine 
MoE with a latent space model and rely on an encoder-decoder architecture to perform distillation for motion learning.  \citet{ranganath2019low} utilize principal component analysis to extract coactivations from reference motions and use them as the atomic actions for motor skill learning.

Despite achieving impressive results, 
exploring the latent space or learning how to combine expert policies is not always easier compared to performing exploration directly in the original action space.
We note that all of these works focus only on reusing models that provide full-body motions.
In contrast, we propose an incremental learning approach that allows a newly trained policy to take only partial actions from a pre-trained policy,
and add on that to generate composite motions.
Our approach can largely reduce the training time for composite and multi-objective tasks involving multiple imitation and goal-directed objectives as 
compared to training from scratch.

\subsection{Multi-Objective Control}
In multi-objective character control, the reward function of the underlying optimization problem is expressed as the weighted sum of multiple, possibly competing, goals. Depending on the task in hand, we seek for objective terms that encourage the character to accomplish behavior goals, follow reference motion and/or style,  
adopt certain behavior characteristics such as low energy movement, attaining specified goals, etc., resulting in an extensive list of objective terms (see~\cite{peng2018deepmimic,abe2007multiobjective,muico2009contact,ye2010optimal,ye2010synthesis,macchietto2009momentum,wu2010goal} for some examples). 
But how we handle all these competing objectives to create coherent, natural, and coordinated control remains an open question. 
A common solution is to employ a manual weighting scheme based on intuition, experience, and trial and error.  However, such approaches often require excessive, often tricky manual effort to obtain desired results. 
While prioritized-based schemes have been employed that optimize each term in the reward function based on a given priority~\cite{de2009prioritized,de2010feature}, 
such schemes cannot automatically address the problem of multiple competing objectives.

This problem becomes worse within a reinforcement learning setting, as small changes in the reward function can have a significant impact on the resulting behavior. 
It may need laborious work to finetune the weight of each objective to ensure that the control policy can effectively balance the learning of multiple objectives in a desired way. 
For tasks with hierarchical objectives,
hierarchical reinforcement learning with multiple controllers can be employed, where a different controller is selected at different task levels~\cite{peng2017deeploco,clegg2018learning,xie2020allsteps,nachum2019multi}.
However, such approaches cannot work for nonhierarchical tasks, where different objective terms need to simultaneously be optimized such as when the character has to perform composite motion imitation and goal-directed control as in our problem domain. In our approach, we propose the use of a multi-critic optimization scheme, where 
each objective is regarded as an independent task and is assigned a separate critic.
By evaluating each objective independently, 
the contribution (gradient) of each objective can be normalized into the same scale,
and, thus, the control policy will be updated toward each objective at the same pace.
As such, we avoid scalarizing and weighting the rewards or priorities of multiple objectives.
In addition, our approach provides a simple solution to adaptively balance the multiple objectives during policy updating without needing to find or estimate the Pareto front.

\begin{figure*}[t]
    \centering
    \includegraphics[width=.775\linewidth]{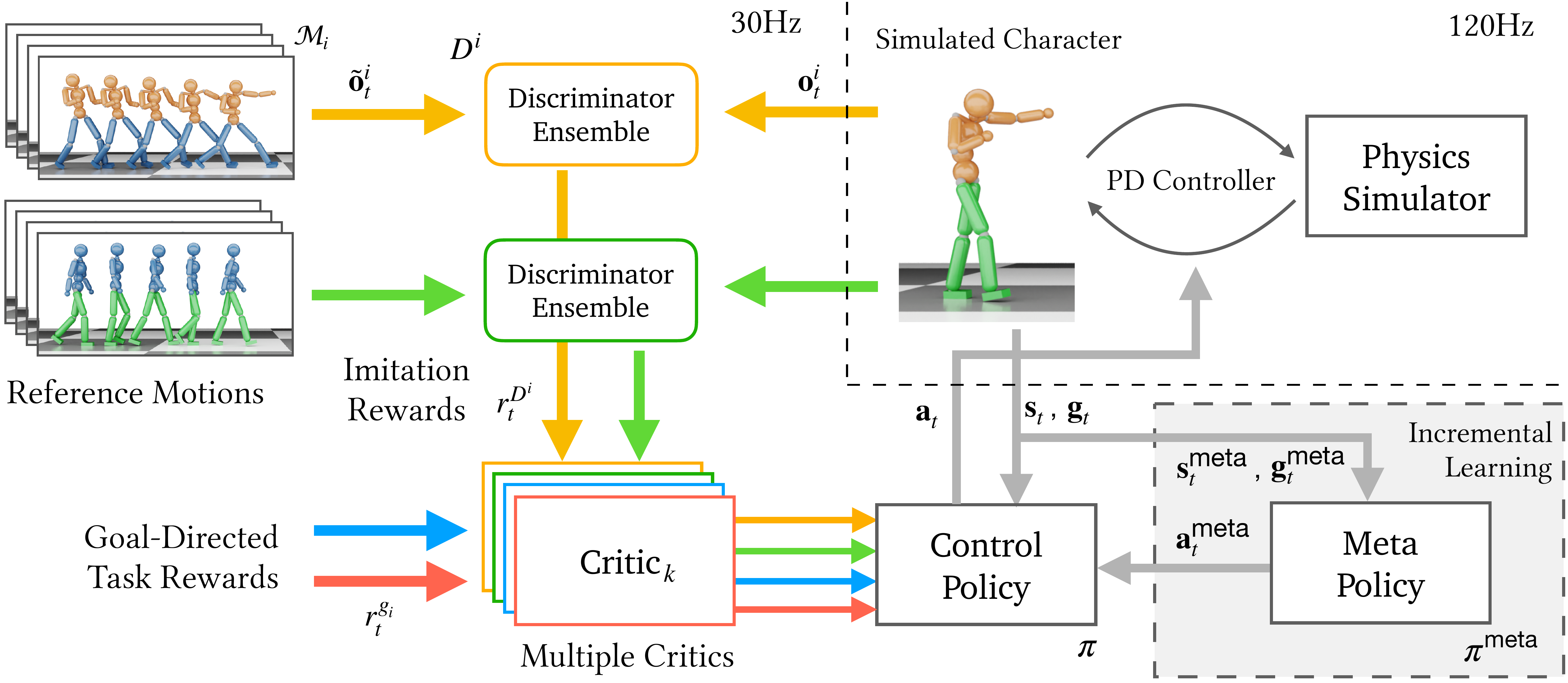}
    \caption{Overview of the proposed system for composite motion learning with task control. 
    Under the framework of reinforcement learning combined with a GAN-like structure for motion imitation,
    our approach employs a multi-critic architecture to train a physics-based controller 
    involving multiple objectives.
    Based on this system,
    we further propose an optional incremental learning scheme that allows the control policy to fast learn new composite motions and tasks by reusing a pre-trained, meta policy.}
    \label{fig:overview}
\end{figure*}

\section{Overview}
Our approach enables a physically simulated character to perform composite motions
through imitating partial-body motions from multiple reference sources directly and simultaneously.
This scheme turns the full-body motion imitation task into a multi-objective optimization problem, 
to which we can further introduce extra objectives for goal-directed control. 
We refer to Fig.~\ref{fig:overview} 
for an overview of our proposed system for composite motion learning with task control. 
We employ a GAN-like structure combined with reinforcement learning to train the control policy imitating the given reference motions.
As such, we do not have to manually design a reward function for imitation learning or explicitly track a target pose from the reference motions.
To learn composite motions, 
we decouple the full-body motion into several partial-body groups each of which imitates its own references. 
Based on this GAN-like structure, 
we propose a multi-objective learning framework that 
exploits multiple critics at the same time to help the control policy learn from multiple objectives, involving both composite motion imitation and goal-directed task control in a balanced way (Section~\ref{sec:composite}). 
To accelerate training, we further consider an optional incremental learning scheme
that reuses a pre-trained policy as the meta policy and 
allows a cooperative policy to adapt the meta one for new composite tasks (Section~\ref{sec:inc_learning}).

\section{Composite Motion Learning}
\label{sec:composite}
Given a physically simulated character, 
we seek 
to train a control policy $\pi(\mathbf{a}_t \vert \mathbf{s}_t, \mathbf{g}_t)$ that simultaneously imitates motions from multiple reference ones, each focusing on specific body parts, 
while possibly completing specific goal tasks. 
At each time step $t$, the control policy takes the character state $\mathbf{s}_t$ and a dynamic goal state variable $\mathbf{g}_t$ as the input
and outputs the control signal (action) $\mathbf{a}_t$.
We let $\mathbf{g}_t$ be an empty variable if no goal-directed control is involved. 
In the following, we detail our proposed approach for training $\pi$ that decouples full-body motion allowing imitation performance to be evaluated and improved with respect to specific body parts, and converts the underlying  composite motion learning problem into a multi-objective optimization problem. 

\subsection{Full-Body Motion Decoupling}\label{sec:motion_decoupling}
At each time step $t$, we represent the character pose 
as $\mathcal{P}_t := \{(p_l, q_l, \dot{p}_l, \dot{q}_l) \vert t\}_{l=1}^{N_\text{link}}$, 
where $p_l \in \mathbb{R}^3$ and $q_l \in \mathbb{R}^4$ are the position and orientation (measured in the unit of quaternion) of each body link respectively, and $\dot{p}_l \in \mathbb{R}^3$ and $\dot{q}_l \in \mathbb{R}^3$ are the linear and angular velocities respectively.
Given the geometry model and joint constraints of the simulated character, this representation can be converted into a joint space one defined by the skeletal joints' local position and velocity and the root's global position and orientation.

Let $\mathcal{M} \supset \{\mathcal{\tilde{P}}_t\}_t $ be the collection of reference motions which may contain multiple clips of pose trajectories $\{\mathcal{\tilde{P}}_t\}_t$ as the reference.
To perform imitation learning,
existing approaches either use a carefully designed reward function to compute the error between $\mathcal{P}_{t+1}$ and $\mathcal{\tilde{P}}_{t+1}$~\cite{peng2018deepmimic,park2019learning,bergamin2019drecon,nuttapong2018,ScaDiver}, 
or employ an evaluator to assess the transfer $\mathcal{P}_t \rightarrow \mathcal{P}_{t+1}$ without explicitly comparing to any specific poses in the reference motions~\cite{merel2017learning,iccgan2021,peng2021amp}.
The former approaches usually need a motion tracking or generation mechanism to retrieve $\mathcal{\tilde{P}}_{t+1}$ from the reference motions.
The latter typically build on the framework of adversarial generative networks (GANs) and rely on a discriminator to evaluate the transfer.
Some approaches take poses from more than one frame during imitation performance evaluation in order to apply more constraints on the pose trajectory.
Nevertheless, all these approaches leverage the full-body character pose $\mathcal{P}_t$ 
and reference pose
$\mathcal{\tilde{P}}_t \in \mathcal{M}$ to perform imitation learning, and thus
intend to learn the full-body motions in $\mathcal{M}$. 

To learn composite motions, 
ideally, we want the simulated character's partial body motions to come from different reference sources at a given time step $t$, i.e., the transfer of pose trajectory $\mathcal{P}_{t-n_i:t}^i \rightarrow \mathcal{P}_{t+1}^i$ 
should satisfy
\begin{equation}\label{eq:partial_imitate}
    \{\mathcal{P}_{t-n_i}^i, \cdots, \mathcal{P}_t^i, \mathcal{P}_{t+1}^i\} \subset \mathcal{M}^i, 
\end{equation}
where $\mathcal{P}_t^i \subset \mathcal{P}_t$ is a partial-body pose from the simulated character, and $\mathcal{M}^i \supset \{\mathcal{\tilde{P}}_t^i\}_t$ is the reference motion collection containing only poses of the partial body group $i$.
The full-body motion is constrained by using multiple $\mathcal{M}^i$ at the same time.
Here, we follow~\citet{iccgan2021} 
and use a pose trajectory having $n_i+2$ frames for imitation performance evaluation.
The larger $n_i$ is, the stricter the evaluation will be, as an error occurring at an earlier time step would negatively influence the evaluation of the following steps. 

Typical partial body groups for a humanoid character would be the upper and lower body, arms, and torso.
For example, we can let $\mathcal{M}^{\text{upper}}$ be a collection of greeting motions involving the upper body (arms, hands, torso and head), and $\mathcal{M}^{\text{lower}}$ be 
walking motions involving the lower body (pelvis, legs and feet).
Then, the full body motion is expected to be the composite of $\mathcal{M}^{\text{upper}}$ (greeting) and $\mathcal{M}^{\text{lower}}$ (walking).
To coordinate the motions from multiple body groups, we can let $\mathcal{P}_t^i$ and some other partial-body poses $\mathcal{P}_t^j$ share some common body link states.
For example, let $\mathcal{P}_t^{\text{upper}}$ and $\mathcal{P}_t^{\text{lower}}$ share the state of one leg to avoid ipsilateral walking.
Correspondingly, the leg state should be included in both $\mathcal{M}^{\text{upper}}$ and $\mathcal{M}^{\text{lower}}$ for the control policy to learn. 
We refer to Sections~\ref{sec:experiments} and~\ref{sec:limitations} for body splitting schemes used in our experiments, including typical upper and lower body decoupling schemes and more tailored ones for specific tasks such as juggling while walking. 
After decoupling the character's full-body motion into multiple sets of $\{\mathcal{P}_t^i\}_t$,
we perform imitation learning with respect to each body group independently, where
the control policy is expected to explore how to combine partial-body motions by itself without needing any full-body, composite motions to be provided as the reference.

\subsection{Imitation Learning}\label{sec:imitation_learning}
To perform imitation learning, 
we build our approach off of GAN-like 
frameworks~\cite{ho2016generative,merel2017learning}, which utilize a discriminator to evaluate imitation performance and generate reward signals for policy optimization using reinforcement learning algorithms.
However, instead of using only one discriminator to perform full-body imitation performance evaluation,
we employ multiple discriminators simultaneously, each of which deals with a body part group $i$ associated with a collection of partial-body reference motions $\mathcal{M}^i$.
Based on this framework, we can avoid designing reward functions to compute the imitation error for each specific body part group.
Furthermore, each discriminator can take only its own interested body link states as input during training. 
Therefore, the provided $\mathcal{M}^i$ 
can still be a collection of full-body motions, but  there is no 
need to explicitly generate any partial-body motions during preprocessing.

To stabilize the adversarial training process, 
we introduce a hinge loss~\cite{lim2017geometric}, gradient penalty term~\cite{gulrajani2017improved}, and an ensemble technique for training of discriminators as proposed in~\cite{iccgan2021}.
Following the literature,
given $\mathbf{o}_t^{i}$ as the observation sampled from the simulated character and $\mathbf{\tilde{o}}_t^{i}$ as that sampled from the reference motions $\mathcal{M}^i$,
the $i$-th ensemble of $N$ discriminators, $D^i = \{D_n^i \vert n = 1, \cdots, N\}$
is trained using the loss function:
\begin{equation}\label{eq:dis_loss}\begin{split}
    \mathcal{L}_{D^i} = \frac{1}{N}\sum_{n=1}^{N}\Bigl( \mathrm{E}_t\left[\max(0, 1+D_n^i(\mathbf{o}_t^{i}))\right] + \mathrm{E}_t\left[\max(0, 1-D_n^i(\mathbf{\tilde{o}}_t^i))\right] \\
    + \lambda^{\text{GP}} \mathrm{E}_t\left[(\vert\vert \nabla_{\mathbf{\hat{o}}_t^i} D_n^i(\mathbf{\hat{o}}_t^i) \vert\vert_2 - 1)^2\right]\Bigr) \qquad
\end{split}\end{equation}
where $\mathbf{\hat{o}}_t^i = \alpha \mathbf{o}_t^i + (1-\alpha) \mathbf{\tilde{o}}_t^i$ with $\alpha \sim \textsc{Uniform}(0, 1)$
and $\lambda^{\text{GP}}$ is gradient penalty coefficient. 

According to Eq.~\ref{eq:partial_imitate}, we define the observation space of a discriminator as
\begin{equation}\label{eq:ob}
    \mathbf{o}_t^i := \{\mathcal{P}_{t-n_i}^i, \cdots, \mathcal{P}_t^i, \mathcal{P}_{t+1}^i\}.
\end{equation}
In principle, the discriminator relies on $\mathbf{o}_t^i$ to evaluate the control policy's performance during the state-action-state transition $(\mathbf{s}_t, \mathbf{a}_t, \mathbf{s}_{t+1})$.
The observation space theoretically should satisfy
$\mathbf{o}_t^i \subseteq \{\mathbf{s}_t, \mathbf{s}_{t+1}\}$.
Otherwise, the discriminator may rely on features unknown to the control policy, and thus it cannot effectively evaluate the policy's performance.
Given that the control policy $\pi$ in our 
formulation 
is still a full-body control policy, we simply define $\mathbf{s}_t$ as a full-body motion state:
\begin{equation}\label{eq:s}
    \mathbf{s}_t := \{\mathcal{P}_{t-n}, \cdots, \mathcal{P}_t\}
\end{equation}
where $n \geq n_i$ for all $i$.
We refer to the Appendix in the supplementary material for more details about the state and observation representation.

The hinge loss function provides a linear evaluation between $[-1, 1]$ to measure the similarity of a given pose trajectory sample $\mathbf{o}_t^{i}$ to any sample in the reference motions.
Therefore, we define the reward term that evaluates the policy's imitation performance 
with respect to $\mathcal{M}^{i}$ for the body part group $i$ at time $t$ as:
\begin{equation}\label{eq:dis_rew}
    r_t^{D^i}(\mathbf{s}_t, \mathbf{a}_t, \mathbf{s}_{t+1}) = \frac{1}{N} \sum_{n=1}^N \textsc{Clip}\left(D_n^i(\mathbf{o}_t^{i}), -1, 1\right).
\end{equation}
It must be noted that even though $\mathbf{o}_t^{i}$ and $\mathbf{\tilde{o}}_t^i$ in Eq.~\ref{eq:dis_loss} have the same subscript $t$, they are paired only for the gradient penalty computation (last term in Eq.~\ref{eq:dis_loss}).
The discriminator ensemble here only evaluates the pose trajectory $\mathbf{o}_t^{i}$ independently,
rather than comparing it against any specific target trajectory.
Therefore, $\mathbf{\tilde{o}}_t^i$ can be randomly sampled from the reference motions by interpolation. 

Overall, by employing multiple discriminator ensembles at each time step $t$, 
we will have a set of rewards, $\{r_t^{D^i}\}_{D^i}$,
to evaluate the policy's performance of controlling the character to perform composite motions, 
i.e. simultaneously imitating different sets of reference motions corresponding to specific partial body parts. 
By doing so, we convert the task of composite motion learning to a multi-objective optimization problem under the framework of reinforcement learning.

\subsection{Multi-Objective Learning}\label{sec:mo_learning}
We consider policy optimization 
of a typical on-policy policy gradient algorithm by maximizing 
\begin{equation}\label{eq:policy_gradient}
    \mathcal{L}_\pi = \mathrm{E}_t[A_t \log \pi(\mathbf{a}_t \vert \mathbf{s}_t, \mathbf{g}_t)], 
\end{equation}
where $\mathbf{s}_t$ and $\mathbf{g}_t$ are the given character's and goals' state variables respectively, and $A_t$ is the advantage which is typically estimated by $\{r_\tau\}_{\tau \geq t}$.
In the common actor-critic architecture, a separate network (critic) is updated in tandem with the policy network (actor).
The critic is employed to provide state-dependent value estimation,
$V(\mathbf{s}_t) = \mathrm{E}_\pi[\sum_{\tau \geq t} \gamma^{\tau-t} r_\tau] = \mathrm{E}_\pi[r_t + \gamma V(\mathbf{s}_{t+1})]$,
based on which $A_t$ can be estimated with less variance,
where $\gamma$ is the discount factor regulating the importance of the contribution from future steps. 
To stabilize the training, standardization is often applied on $A_t$ where the standardized advantage $\bar{A}_t$ is used in place of $A_t$ for policy updating.   

A typical solution for multi-objective tasks in reinforcement learning is to simply add together all objective-related reward terms, $r_t^k$, with some weights $\omega_k$, i.e., $r_t = \sum_{k=1}^K \omega_k r_t^k$ for a $K$-objective problem.
In such a way, we still have a scalar reward that can be used with Eq.~\ref{eq:policy_gradient} for policy updating. 
In practice, though, given that conflicts may exist among the different reward terms,
manually tuning the values of $\omega_k$ to balance the composite objective of the character is not an intuitive task. 
For example, we may need the policy to put more effort into learning a difficult partial-body motion, instead of  
even with a trade-off in learning other motions, 
rather than only focusing on the easy ones to keep achieving a higher associated reward.
In addition,  
our proposed approach performs reward estimation by 
employing multiple discriminators simultaneously, which are modeled by neural networks. 
This scheme brings a lot of uncertainty,  
as the reward distributions from different discriminators may differ a lot depending on the given reference motions, which could be unpredictable before training.
Such a problem would deteriorate if we further introduce a set of goal-directed tasks, each having its own associated reward term which may compete against the imitation reward terms. 

To balance the contributions of multiple objectives during policy updating,
we propose to model the multi-objective learning problem as a multi-task one,
where each objective is taken into account as an \emph{independent task} and has a fixed importance during policy updating.
To do so,
instead of using $r_t = \sum_k \omega_i r_t^k$,
we compute the advantage of $A_t^k$ with respect to $\{r_\tau^k\}_{\tau \geq t}$ independently.
Then, the optimization process becomes maximizing
\begin{equation}\label{eq:policy_loss}
    \mathcal{L}_{\pi} = \sum_{k=1}^K \mathrm{E}_t\left[\omega_k \bar{A}_t^k \log \pi(\mathbf{a}_t \vert \mathbf{s}_t, \mathbf{g}_t)\right], 
\end{equation}
where $\sum_k \omega_k = 1$ and $\bar{A}_t^k$ is the standardization of $A_t^k$, i.e.
\begin{equation}\label{eq:adv_normalize}
    \bar{A}_t^k = \frac{A_t^k - \mathrm{E}_t[A_t^k]}{\sqrt{\mathrm{Var}_t[A_t^k]}}.
\end{equation}
This optimization process is equal to updating the policy with respect to each objective independently but always at the same scale proportional to $\omega_k$.
The introduction of $\omega_k$ gives us more flexibility to adjust the contributions toward each objective when conflicts occur during policy updating.
However, under our testing,
a simple choice of $\omega_k = 1/K$, which means each objective is equally important, works well for most cases.
We refer to the Appendix in the supplementary material for the choice of $\omega_k$ in our tested composite tasks.

During implementation, we can rewrite Eq.~\ref{eq:policy_loss} as
\begin{equation}\label{eq:policy_loss_one_pass}
    \mathcal{L}_{\pi} = \mathrm{E}_t\left[\left(\sum\nolimits_k \omega_k \bar{A}_t^k\right) \log \pi(\mathbf{a}_t \vert \mathbf{s}_t, \mathbf{g}_t)\right]
\end{equation}
such that the policy update can be done through backward propagation in one pass.
From this equation, we can see that the nature of our approach is to introduce a dynamic coefficient constrained by the standard deviation of $\{A_t^k\}_t$ for each objective $k$.
As such, the policy will be updated with respect to each objective adaptively. 
This separation of objectives leads to a single-policy multi-critic architecture. 
In Fig.~\ref{fig:overview}, for example,
we have two imitation related reward terms (yellow and green) for upper and lower body imitation respectively, and two goal-directed task reward terms (red and blue).
Accordingly, we employ four critics denoted by $\textsc{Critic}_k$ in the figure.
Each $\textsc{Critic}_k$ only participates in the estimation of $A_t^k$, and takes the reward associated with the objective $k$, i.e. $\{r_t^k\}_t$, for training.

Though the policy update is balanced through the 
proposed multi-critic architecture,
the state values, which are decided by $\{r_t^k\}_t$,  could differ still drastically with respect to each objective 
depending on the difficulty of given reference motions or the reward distributions of the goal-related tasks.
To mitigate this issue and stabilize the training of critics,
we introduce the value normalization scheme of \textit{PopArt}~\cite{van2016learning}.
The value target under this scheme is normalized by the moving average and standard deviation for the critic network training.
The output of a critic is unnormalized before joining the process of advantage estimation.
Besides maintaining a normalizer for value targets,
\textit{PopArt} is designed to preserve the output precisely.
Namely, with \textit{PopArt}, the output of a critic is identical before and after the normalizer updates given the same input state $\mathbf{s}_t$ and $\mathbf{g}_t$.
Such a design is to prevent the normalization from affecting the value state estimation, thereby stabilizing the policy training.
In our implementation, each critic $\textsc{Critic}_k(\mathbf{s}_t, \mathbf{g}_t)$ has its own normalizer with a scalar scale and shift estimated independently with respect to its associated objective $k$.
As we show in Section~\ref{sec:ablation}, the introduction of \textit{PopArt} 
helps improve the policy performance as also demonstrated by previous works~\cite{yu2021surprising,van2016learning}.  

\begin{algorithm}[t]
Prepare the meta policy $\pi^{\text{meta}}$\;
initialize the policy network $\pi$\;
initialize the critic network $\textsc{Critic}_k$ where $k = 1, \cdots, K$ given $K$ objectives in the task\;
initialize policy replay buffer $\mathcal{T}$ and reward buffer $\mathcal{R}$\;
prepare reference motions $\mathcal{M}^i$ for each discriminator ensemble $D^i$\;
\While{training does not converge}{
    $\mathcal{T} \gets \emptyset$, $\mathcal{R} \gets \emptyset$\;
    \For{each environment step $t$}{
        $\mathbf{a}_t^{\text{meta}} \sim \pi^{\text{meta}}(\cdot \vert \mathbf{s}_t^{\text{meta}}, \mathbf{g}_t^{\text{meta}})$\;
        $\mathbf{a}_t \sim \pi(\cdot, \vert \mathbf{s}_t, \mathbf{g}_t, \mathbf{a}_t^{\text{meta}})$\;
        $\mathbf{s}_{t+1}, \mathbf{g}_{t+1}, \mathbf{r}_t^{\mathbf{g}_t} \gets$ environment updates with character control signal of $\mathbf{a}_t$\;

        extract observation $\mathbf{o}_t^{i}$ from the state pair of $\mathbf{s}_t$ and $\mathbf{s}_{t+1}$ for each discriminator ensemble $D^i$\;
        
        $\mathcal{T} \gets \mathcal{T} \cup \{(\mathbf{s}_t, \mathbf{a}_t, \{\mathbf{o}_t^{i}\}_i)\}$\;
        $\mathcal{R} \gets \mathcal{R} \cup \{r_{t+1}^{k}\}$ for each term $k$ in $\mathbf{r}_i^{\mathbf{g}_t}$\;
        $\mathbf{s}_{t} \gets \mathbf{s}_{t+1}$; $\mathbf{g}_{t} \gets \mathbf{g}_{t+1}$\;
        extract $\mathbf{s}_{t}^{\text{meta}}$ and $\mathbf{g}_t^{\text{meta}}$ from $\mathbf{s}_t$ and $\mathbf{g}_t$ respectively\
    }
    \For{each discriminator ensemble $D^i$}{
        draw samples $\mathbf{\tilde o}_t^{D^i}$ from $\mathcal{M}^i$\;
        update $D^i$ using $\mathbf{o}_t^{i}$ from $\mathcal{T}$ and $\mathbf{\tilde o}_t^i$ based on Eq.~\ref{eq:dis_loss}\;
        \For{each $\mathbf{o}_t^{i}$ in $\mathcal{T}$}{
           compute step-wise imitation reward $r_t^{D^i}$ based on Eq.~\ref{eq:dis_rew}\;
           $\mathcal{R} \gets \mathcal{R} \cup \{r_t^{D^i}\}$\
        }   
    }
    \For{each reward term collection $\{r_t^{k}\}_t$ in $\mathcal{R}$}{
        compute advantage $A_t^{k}$ using $\{r_{\tau}^{k}\}_{\tau\geq t}$ and state value estimation from $\textsc{Critic}_k(\mathbf{s}_\tau, \mathbf{g}_\tau)$ unnormalized by \textit{PopArt}\;
        compute value target $V_t^k$ based on $A_t^k$\;
        update the normalizer for $\textsc{Critic}_k$ based on $V_t^k$ using \textit{PopArt}\;
        get normalized value target $\bar{V}_t^k$ by \textit{PopArt}\;
        get normalized advantage $\bar{A}_t^{k}$ based on Eq.~\ref{eq:adv_normalize}\
    }
    \For{each policy update step}{
        update $\pi$ using $\{(\mathbf{s}_t, \mathbf{a}_{t}, \{\bar{A}_t^k\}_k)\}_t$ based on Eq.~\ref{eq:policy_loss_one_pass}\;
        update each critic network $\textsc{Critic}_k$ using $\{\bar{V}_t^k\}_t$\
    }
}
\caption{Multi-Objective Incremental Learning}
\label{alg:alg}
\end{algorithm}

\section{Incremental Learning}\label{sec:inc_learning}
Besides being able to perform a range of composite motions, 
humans typically learn such motions in an incremental manner. For example, if we know how to walk, we should be able to quickly learn how to hold our phone while walking. There is no need to relearn walking from scratch. 
Based on this intuition, 
we propose an incremental learning scheme for fast composite motion learning. 
Instead of training a policy completely from scratch,
we reuse a pre-trained policy as a
 meta policy $\pi^\text{meta}$  
that allows the simulated character to perform a basic set of motions (walking in the previous example). 
Given $\pi^\text{meta}$, we train a new policy $\pi$ to cooperate with the meta policy,  
performing new composite motions by action addition (holding a phone + walking).

Formally, let $\pi(\mathbf{a}_t \vert \mathbf{s}_t, \mathbf{g}_t) := \mathcal{N}(\boldsymbol{\mu}_t, \boldsymbol{\sigma}_t^2)$ denote a Gaussian-based policy.
By introducing a meta policy $\pi^\text{meta}$, we define the policy, which is trained to cooperate with $\pi^\text{meta}$ for new composite motions as
\begin{equation}\label{eq:pi_incremental}\begin{split}
    \pi(\mathbf{a}_t \vert \mathbf{s}_t, \mathbf{g}_t, \mathbf{a}_t^{\text{meta}}) := \mathcal{N}\left(\boldsymbol{\mu}_t, \boldsymbol{\sigma}_t^2\right) + \mathbf{w}_t \textsc{Stop}\left(\mathbf{a}_t^{\text{meta}}\right) \\
    = \mathcal{N}\left(\boldsymbol{\mu}_t + \mathbf{w}_t \textsc{Stop}\left(\mathbf{a}_t^{\text{meta}}\right), \boldsymbol{\sigma}_t^2\right),
\end{split}
\end{equation}
where 
the weight vector $\mathbf{w}_t$ has the same dimension with $\mathbf{a}_t^\text{meta}$, and 
$\mathbf{a}_t^{\text{meta}} \sim \pi^{\text{meta}}(\cdot \vert \mathbf{s}_t^{\text{meta}}, \mathbf{g}_t^{\text{meta}})$ is drawn from the meta policy.
$\mathbf{w}_t$ are defined as a set of weights each of which is associated with a DoF in the action space of the meta policy.
In our implementation,
$\mathbf{w}_t$, $\boldsymbol{\mu}_t$ and $\boldsymbol{\sigma}_t$ are obtained by a neural network taking $\mathbf{s}_t$ and $\mathbf{g}_t$ as input, and thus are learnable.
We put a "gradient stop" operator, $\textsc{Stop}(\cdot)$, on $\mathbf{a}_t^\text{meta}$, which means that the meta policy is fixed and will not be updated with $\pi$.

Using this incremental learning scheme, the new, cooperative policy adds its own action to the meta action $\mathbf{a}_t^{\text{meta}}$.
The weight vector $\mathbf{w}_t$ decides the reliance of $\pi$ on the meta policy $\pi^\text{meta}$ with respect to each DoF in the action space.
The bigger an element in $\mathbf{w}_t$ is, the more the cooperative policy relies on the meta policy to control the corresponding DoF.
As such,
$\pi$ is trained incrementally to learn new composite motions by reusing the meta policy partially.
This scheme does not require that $\mathbf{a}_t^{\text{meta}}$ and $\mathbf{a}_t$ must have exactly the same dimension, as we can assume zero values for the missing dimensions in $\mathbf{a}_t^\text{meta}$ or ignore the extra, uninteresting dimensions in $\mathbf{a}_t^\text{meta}$.
Compared to a mixture-of-experts (MoE) model, where the action is obtained by a linear combination of the actions from multiple expert policies, 
our approach focuses on reusing partial-body motions from the meta policy.
It would be very difficult for a MoE model  
to keep, for example, only the lower-body motion of one expert and replace the upper-body motion with that of another expert through a linear combination of the experts' full-body motions.

With the introduction of $\pi^\text{meta}$,
we can replace $\pi(\mathbf{a}_t \vert \mathbf{s}_t, \mathbf{g}_t)$ in Eq.~\ref{eq:policy_loss} with $\pi(\mathbf{a}_t \vert \mathbf{s}_t, \mathbf{g}_t, \mathbf{a}_t^{\text{meta}})$,
and perform composite motion learning with goal-directed control under our proposed multi-objective learning framework.
We refer to Algorithm~\ref{alg:alg} for the outline of the proposed multi-objective learning framework with incremental learning.
To train a composite policy completely from scratch without using incremental learning, 
we can simply ignore $\pi^\text{meta}$ and
use $\pi(\mathbf{a}_t \vert \mathbf{s}_t, \mathbf{g}_t)$ solely in Algorithm~\ref{alg:alg}.

\section{Experiments}
\label{sec:experiments}
In this section,
we experimentally evaluate our approach on multiple challenging composite motion learning tasks.
We show that our approach can effectively let motor control policies learn composite motions from 
multiple reference motions directly 
without manually generating any full-body motion as reference.
Besides evaluating the imitation performance,
we also apply our approach on several goal-directed control tasks combined with composite motion learning from unstructured reference data. 
The results demonstrate that our proposed approach can successfully tackle complex tasks balancing the learning of multiple objectives involving both partial-body motion imitation and goal-directed control. 
Finally, we perform ablation studies on our proposed multi-objective learning framework and  incremental learning scheme.

\subsection{Implementation Details}
We run physics-based simulations using IsaacGym~\cite{makoviychuk2021isaac},
which supports simulation with a large number of instances simultaneously by leveraging GPU.
The simulated humanoid character
has 15 body links and 28 DoFs,
where the hands are fixed with the forearms and are uncontrollable.
In the tasks involving a tennis player, 
we add 3 DoFs on the right wrist joint 
such that the character can control the racket more agilely, 
though the racket is fixed on the right hand.
The simulation runs at 120Hz and the control policy  at 30Hz.
Differing from the previous works that employ a stable PD controller~\cite{tan2011stable} for character control~\cite{peng2018deepmimic,iccgan2021,peng2021amp,ScaDiver,lee2021learning,deepcompliantcontrol,park2019learning,won2022physics}
we employ a normal, linear PD servo 
for faster simulation.

\begin{figure}[t]
    \begin{subfigure}[t]{.3\linewidth}
    \includegraphics[width=\linewidth]{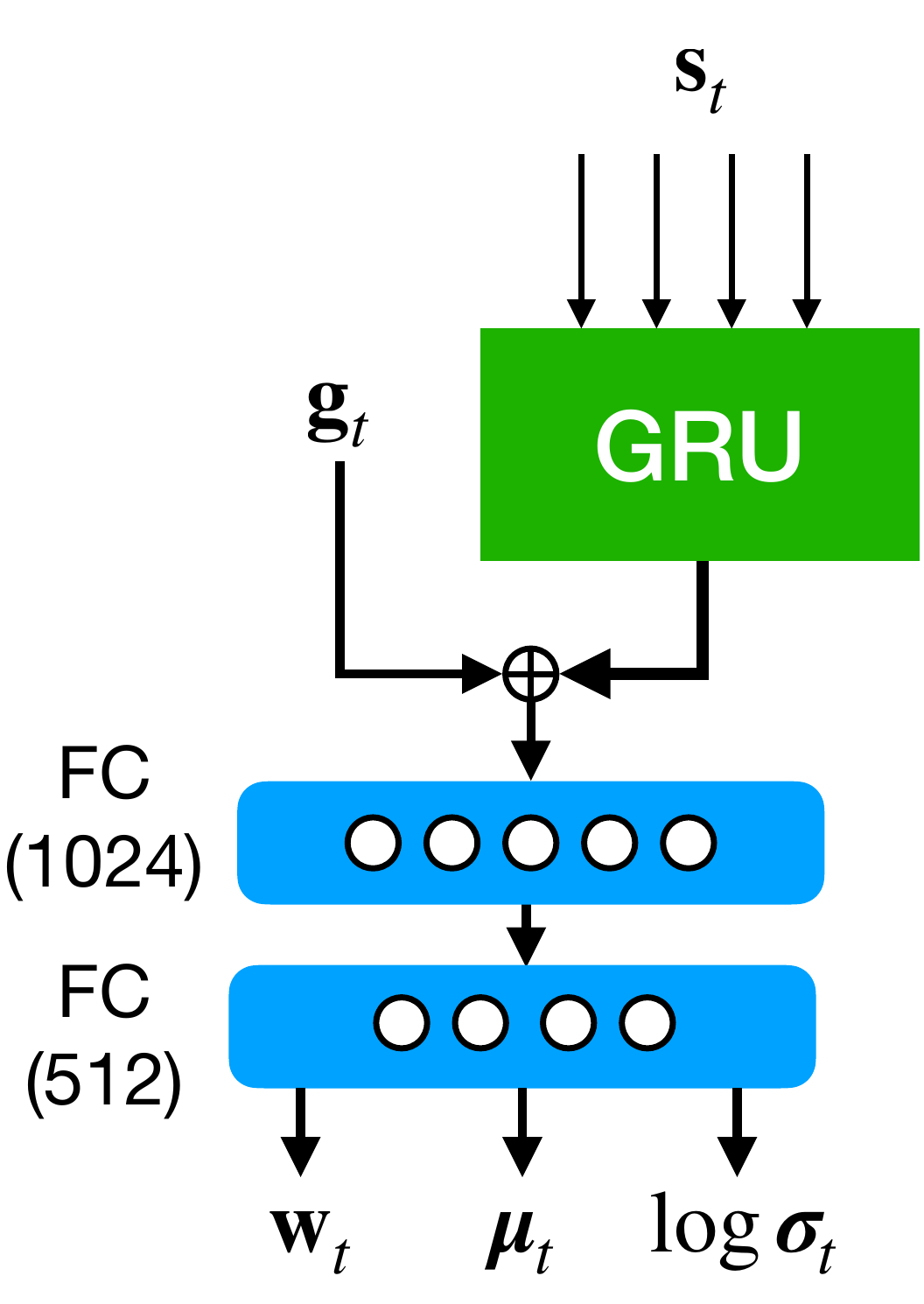}
    \caption{Policy Network}
    \end{subfigure}\hfill
    \begin{subfigure}[t]{.3\linewidth}
    \includegraphics[width=\linewidth]{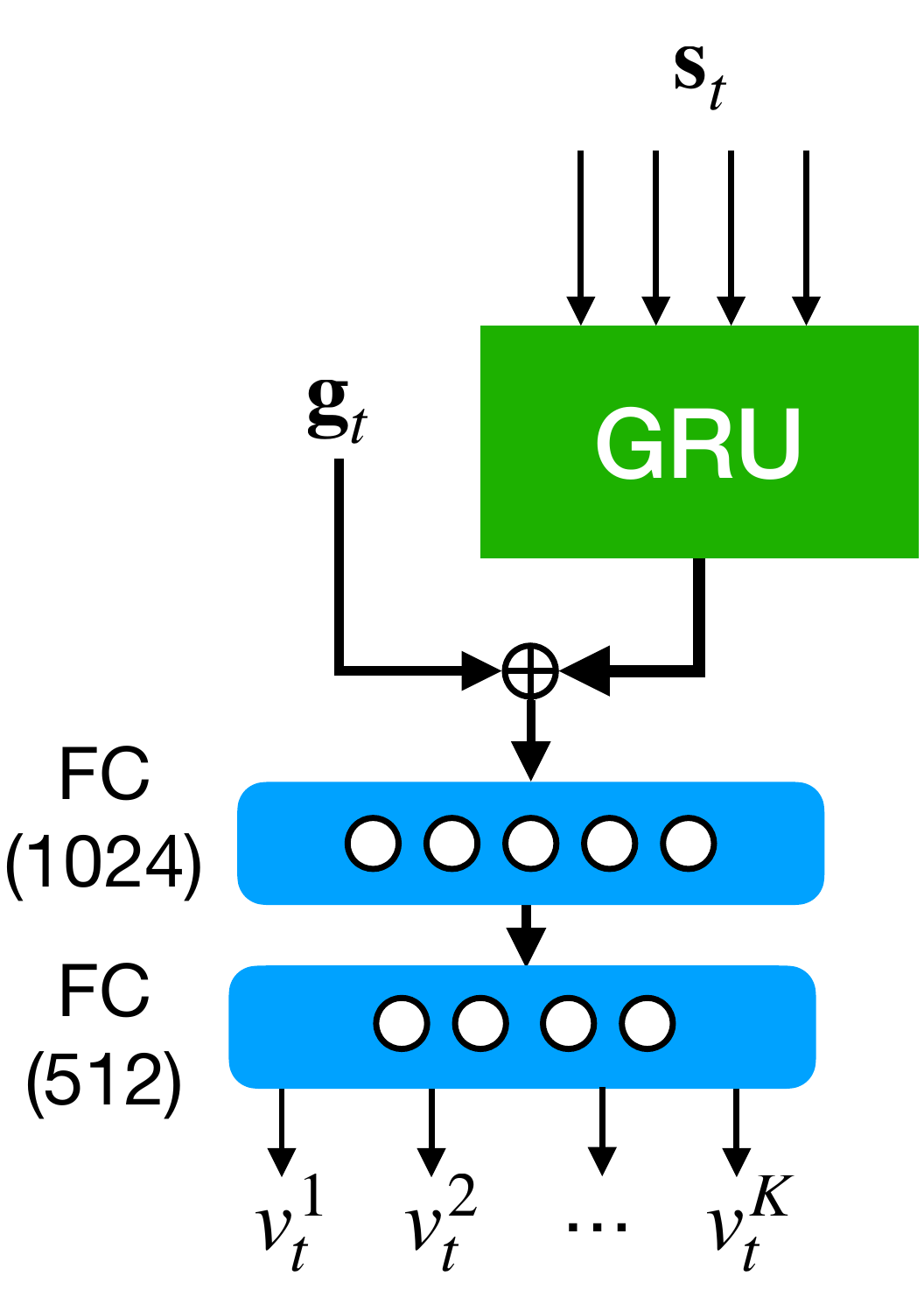}
    \caption{$K$-Head Critic}
    \end{subfigure}\hfill
    \begin{subfigure}[t]{.36\linewidth}
    \centering
    \includegraphics[width=.85\linewidth]{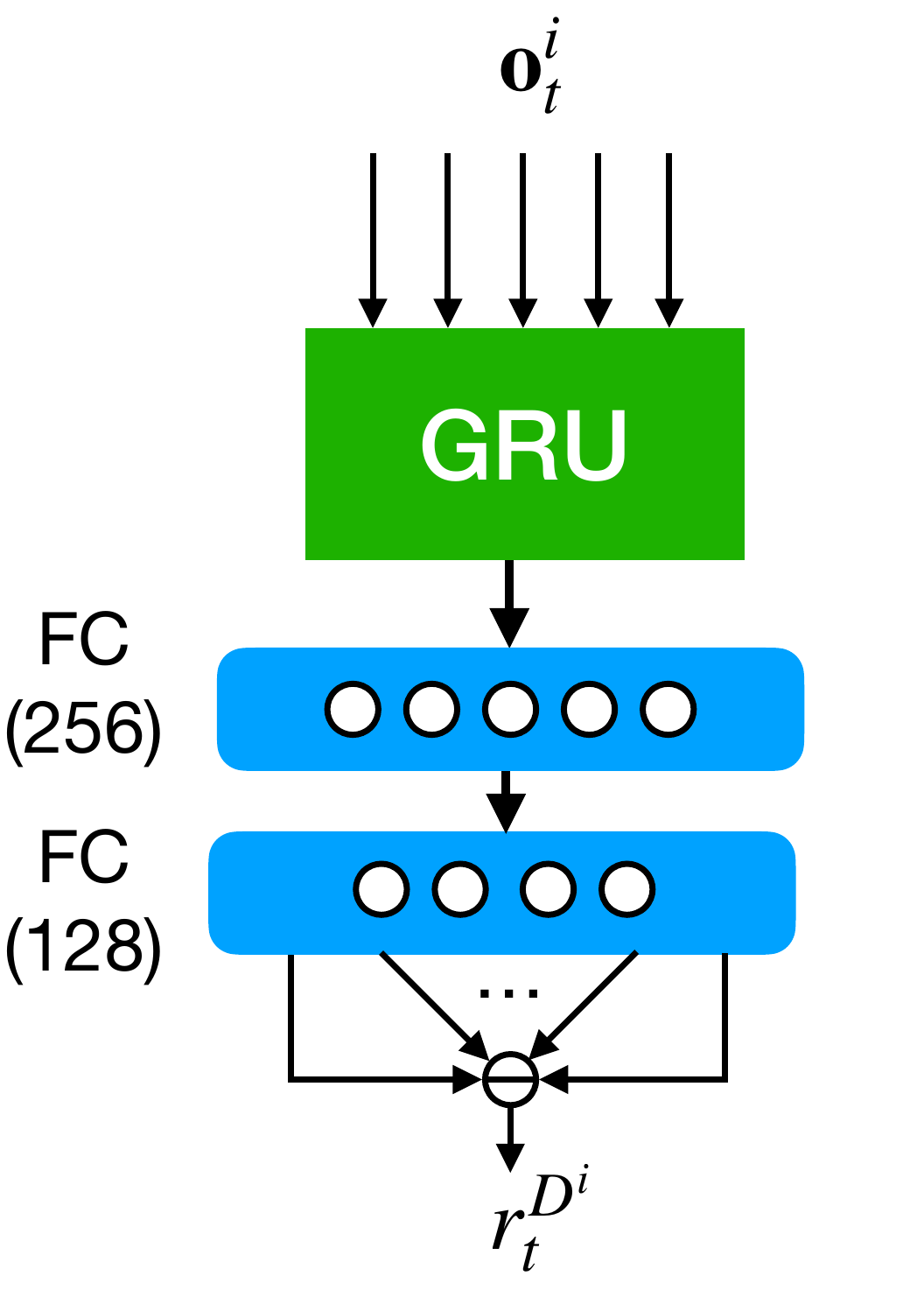}
    \caption{Discriminator Ensemble}
    \end{subfigure}
    \caption{Network structures. $\oplus$ denotes the concatenation operator and $\ominus$ denotes the average operator.}
    \label{fig:network}
\end{figure}
We use PPO~\cite{schulman2017proximal} as the base reinforcement learning algorithm for policy training
and Adam optimizer~\cite{kingma2014adam} to perform policy optimization.  
To embed the character state $\mathbf{s}_t$ and the discriminator observation $\mathbf{o}_t^i$ sequentially,
we employ a gated recurrent unit (GRU)~\cite{chung2014empirical} with a 256-dimension hidden state to process these temporal inputs.
The embedded character state feature is concatenated with the dynamic goal state $\mathbf{g}_t$ if goal-directed control is involved,
and then passed through a multilayer perceptron with two full-connected (FC) layers.
The control policy is constructed as Gaussian distributions with independent components.
The output of the policy network includes the mean $\boldsymbol{\mu}_t$ and standard deviation $\boldsymbol{\sigma}_t$ parameters of the
policy distribution as well as a weight vector $\mathbf{w}_t$ when incremental learning is exploited.
The multiple critics in our multi-objective learning framework are modeled by a multi-head neural network. 
Similarly to the critic networks, we model a discriminator ensemble using a multi-head network. 
The outputs are averaged by Eq.~\ref{eq:dis_rew} to produce the reward signal.
All the network structures are shown in Fig.~\ref{fig:network}, in which we assume that there are $K$ objectives in total. 
We refer to the Appendix in the supplementary material for the representation of $\mathbf{g}_t$ in our designed goal-directed tasks, and all hyperparameters used for policy training.

All the tested policies were trained on a machine equipped with an Nvidia V100 GPU.
It typically takes about 1.5h to train a policy using a fixed budget of 20M samples (environment steps),
for a pure composite motion imitation task.
For complex tasks involving goal-directed control, 
it takes about 15 to 30 hours and requires about $2 \times 10^8$ to $4 \times 10^8$ samples to train a policy from scratch.
By exploiting our incremental learning scheme to reuse a pre-trained meta policy, we can shorten the training time to about 30 minutes to 2 hours depending on the difficulty of the tasks.

\subsection{Data Acquisition}
All the motion data used for training are obtained from 
the LAFAN1 dataset~\cite{harvey2020robust} and other commercial and publicly available motion capture datasets recorded at 30Hz.
For single-clip imitation, 
we synthesize short reference motion clips of 1-3 seconds long (cf. Table~\ref{tab:imit}).
For tasks with goal-directed control,
we extract several collections of motions (cf. Table~\ref{tab:imit_col}), each of which contains multiple clips of reference motions with lengths varying from about 15 to 70 seconds. 
The juggling motion involves a single trial of a subject performing juggling while standing on a skate, while the collection of tennis swing motions contains four trials of forehand swings captured from different subjects. 
We retarget the local joint position from those motion data to our character model without extra manual reprocessing.
We demonstrate that policies trained with our approach can perform motion synthesis from unstructured data for goal-directed control,  
and can explore how to perform composite motions by combining the partial-body motions from the reference motions without needing any manual processing for motion blending.

\subsection{Imitation Performance}\label{sec:imit_perf}

\afterpage{
\begin{figure}[H]
    \centering
    \includegraphics[width=.99\linewidth]{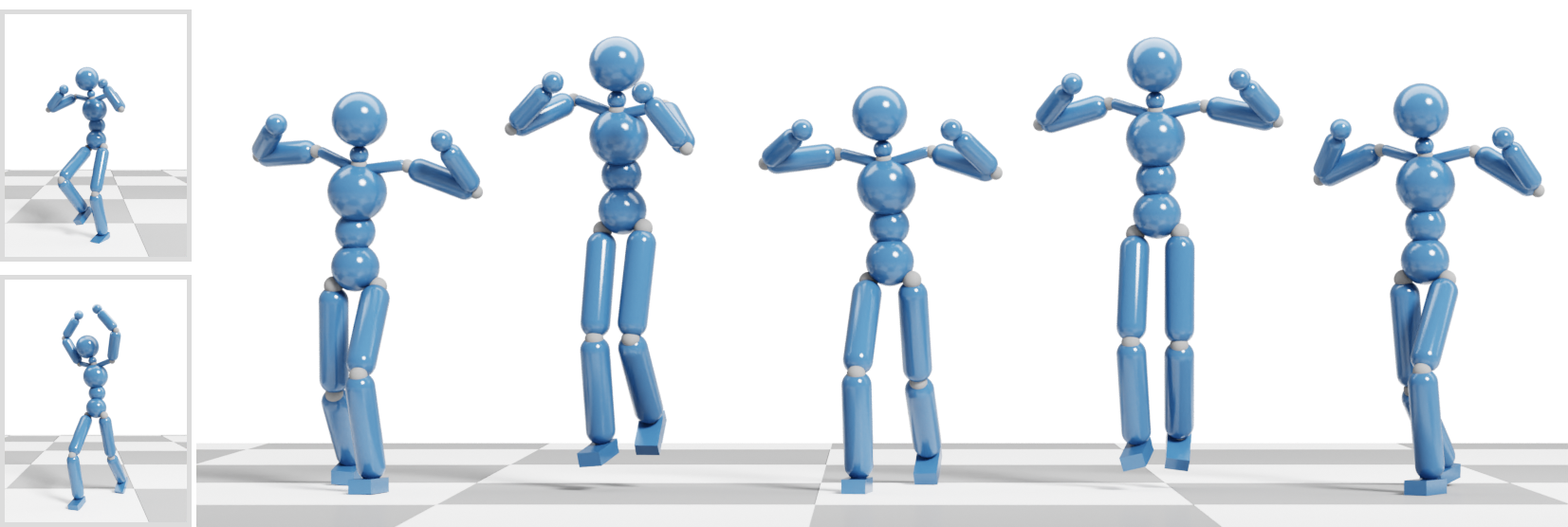}\\\vspace{0.02cm}
    \includegraphics[width=.99\linewidth]{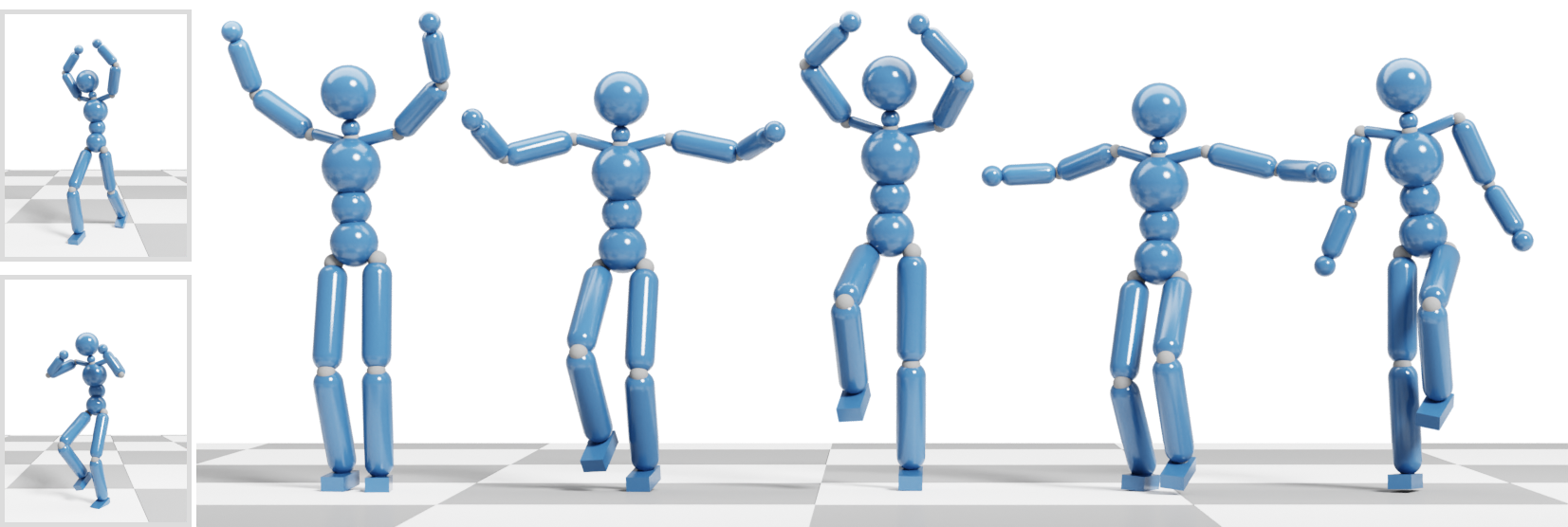}\\\vspace{0.02cm}
    \includegraphics[width=.99\linewidth]{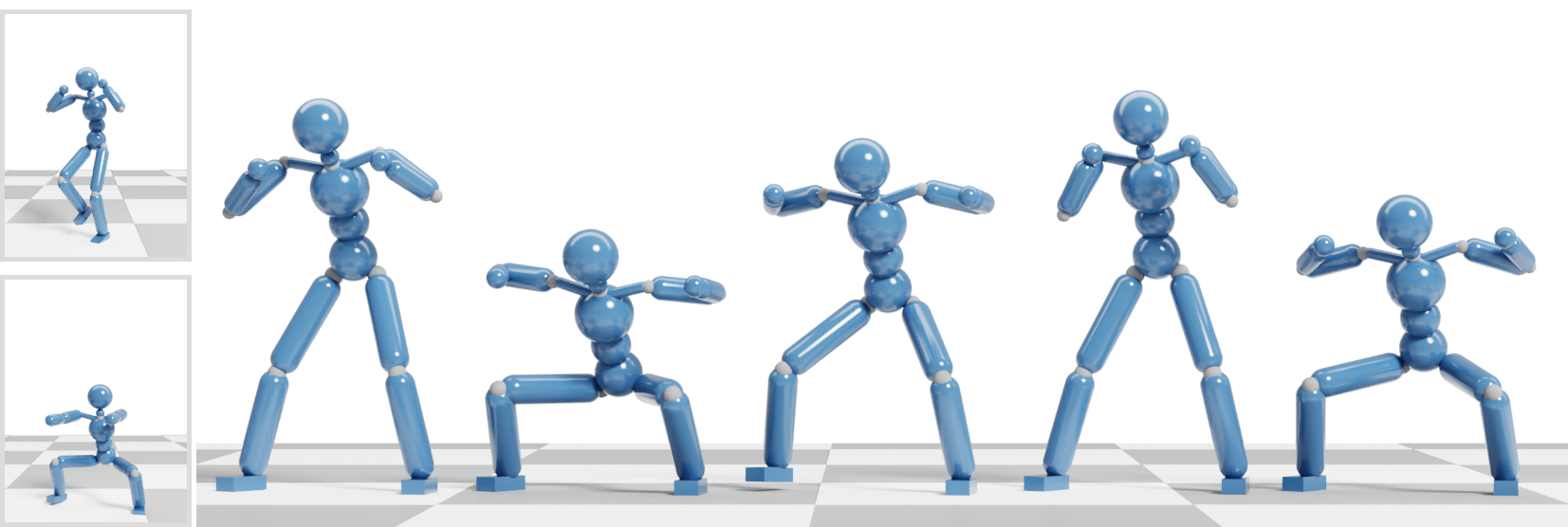}\\\vspace{0.02cm}
    \includegraphics[width=.99\linewidth]{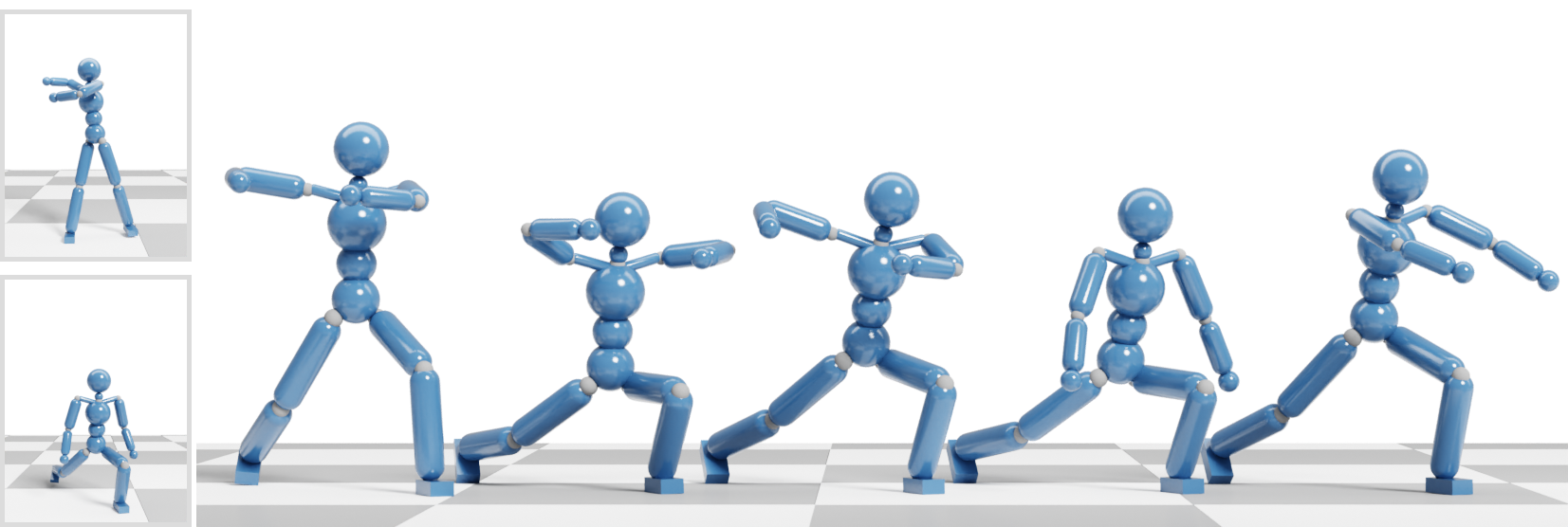}\\\vspace{0.02cm}
    \includegraphics[width=.99\linewidth]{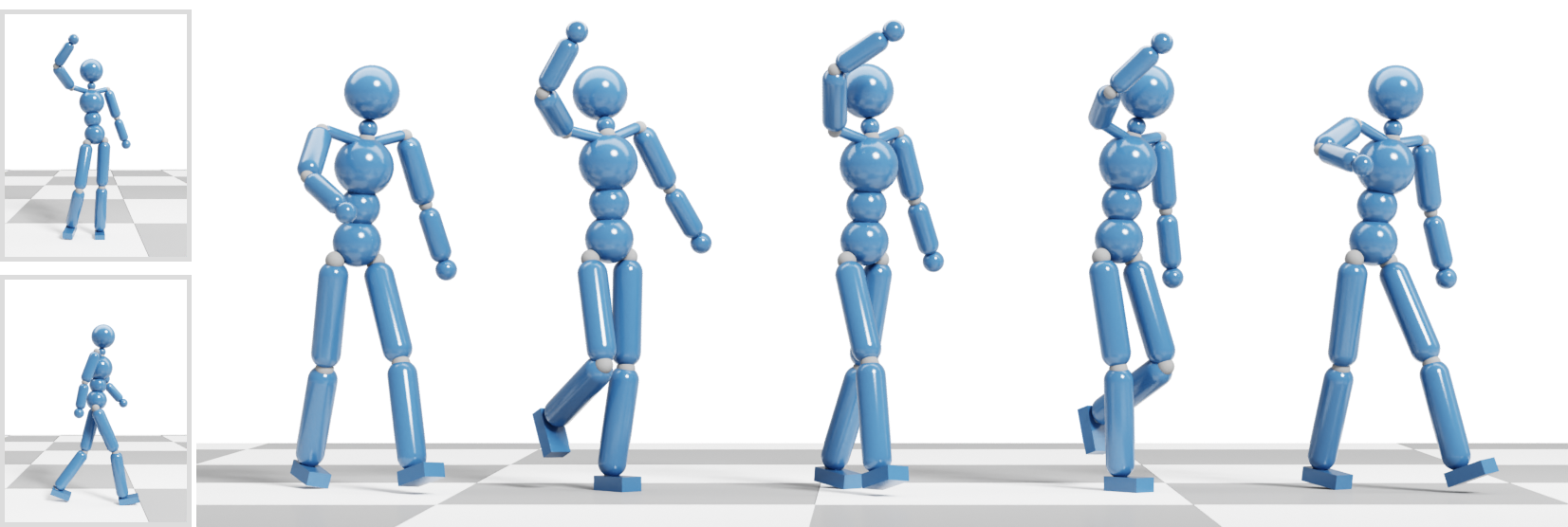}\\\vspace{0.02cm}
    \includegraphics[width=.99\linewidth]{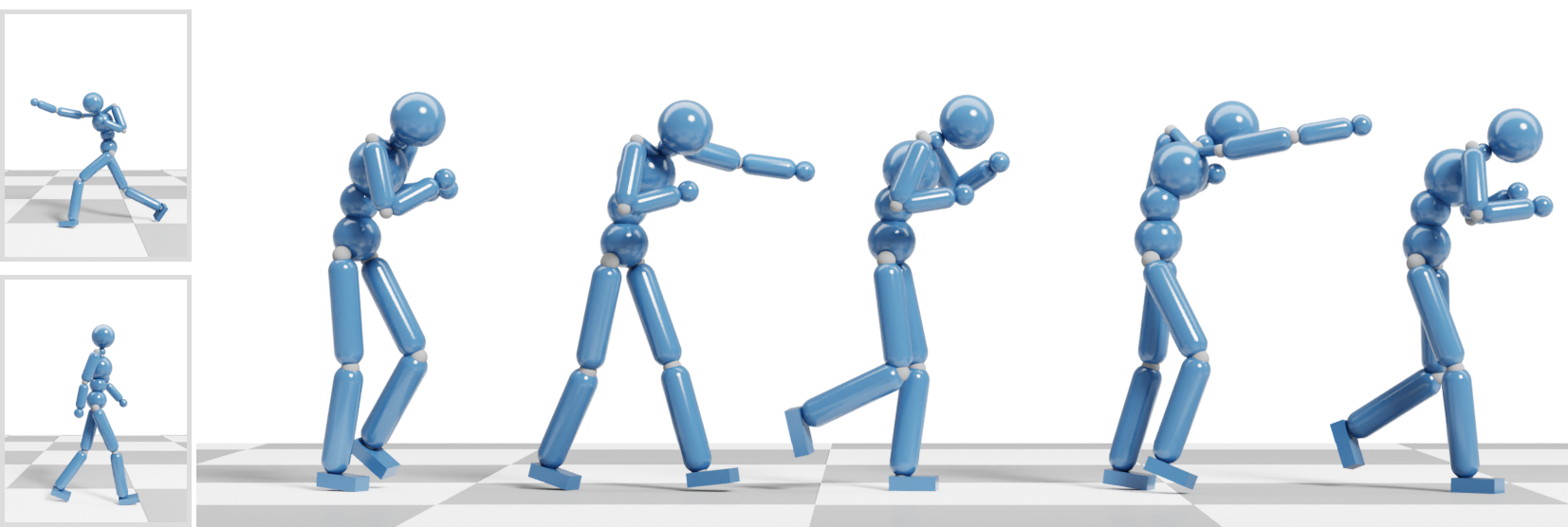}\\\vspace{0.02cm}
    \includegraphics[width=.99\linewidth]{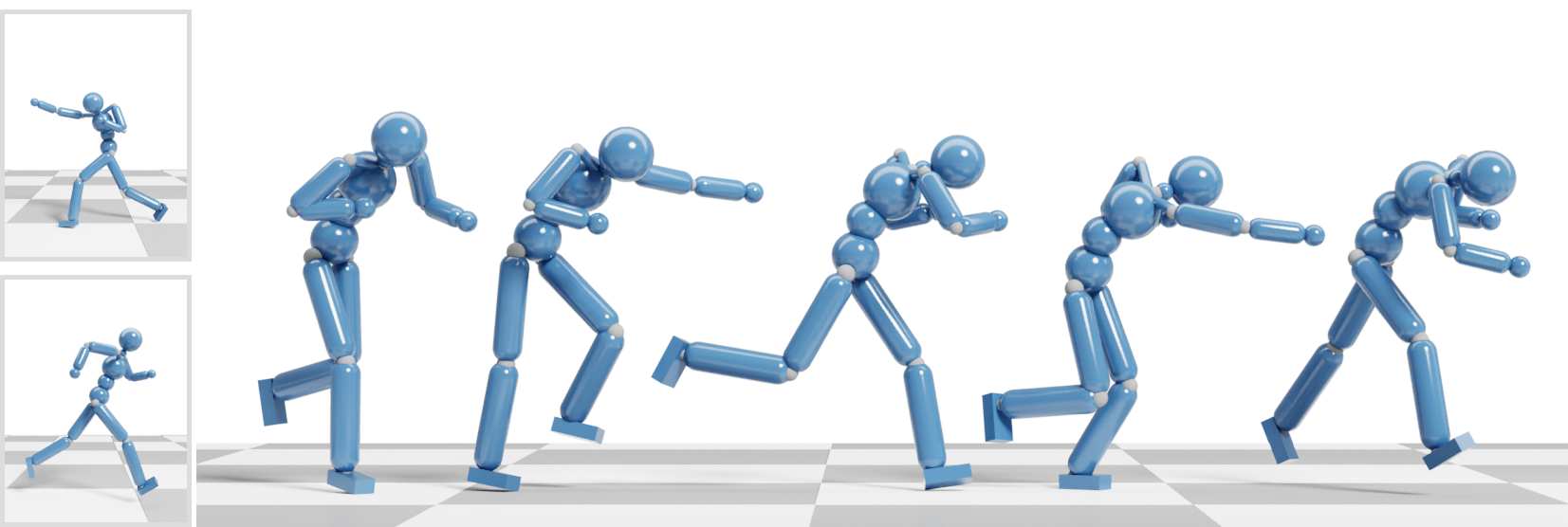}
    \caption{Composite motions learned from multiple single-clip reference motions. 
    The two snapshots shown on the left side of each row are the reference motions for the upper and lower body respectively.}
    \label{fig:comp_motions}
\end{figure}
}

In Fig.~\ref{fig:comp_motions},
we highlight motion pose snapshots captured from some of our trained policies for composite motion learning.
Each composite motion is learned based on two reference motion clips, one 
for the upper body and the other one 
for the lower body.
From top to bottom,
the names of corresponding motions are listed in Table~\ref{tab:imit}.
Overall, 
policies trained with our approach can perform very challenging composite motor skills by using the character's upper and lower body part groups at the same time.
For example, in the motion combination of chest open and jumping jack (1st row), 
the control policy must keep the character's body balanced  to perform the chest-open motion
during jumping in the air,
which is a pretty challenging task even for humans. 
Similar challenges arise when doing squats with the chest open (3rd row) and lunges with waist twisting (4th row).
Besides
simply following the two partial-body reference motions at the same time,
the control policies must master how the partial motions could be combined such that the full-body motion is physically plausible.
In the 4th row,
for example,
it is impossible for the character to keep twisting its waist while doing lunges at quite different frequencies.
Similarly, 
in the motion combination of punch and walk (6th row) and that of punch and run (7th row),
the character's foot has to contact the ground first in order to perform the punch action with the torso leaning forward.
The control policy, thereby, must
know when the punch action is doable and arrange the motion combination by itself,
rather than strictly following the reference motions.
Our approach does not require the given reference motions to be perfectly synchronized.
The control policies take the character state as input and perform composite motions accordingly.
Furthermore, the proposed dynamic sampling rate (see Appendix) allows the control policy to adjust the motion speed within an acceptable range for better motion combining.

To quantitatively evaluate the imitation performance,
following previous literature~\cite{harada2004quantitative,tang2008emulating,peng2021amp,iccgan2021},
we leverage the technique of fast dynamic time warping (DTW) and  measure the imitation error as follows:
\begin{equation}
    e_t = \frac{1}{N_\text{link}^i} \sum_{l=1}^{N_\text{link}^i} \vert\vert p_l - \tilde{p}_l\vert\vert, 
\end{equation}
where 
$N_\text{link}^i = \vert \{\mathcal{P}_t^i\} \vert$ is the number of interesting body links in the $i$-th body part group,
$p_l \in \mathbb{R}^3$ is the position of the body link $l$ in the world space at the time step $t$, and $\tilde{p}_l$ is the body link's position in the reference motion.
The evaluation results are shown in Table~\ref{tab:imit}.
Our approach can imitate the reference motions closely and
balance the imitation of the two partial-body motions well.
As can be seen, there is no big gap between the two imitation errors in a given composite motion combination,
which means that policies trained with our approach do not just follow only one reference motion and ignore the other one. 
In contrast, without using our proposed multi-objective learning framework,
the policy could prefer to track only one reference motion that is easy to follow.
We refer to Section~\ref{sec:ablation} for the related ablation study.

\begin{table}[t]
    \centering\small
    \caption{Imitation performance when learning composite motions from single clips of reference motions.}
    \begin{tabular}{r|cc}
        \toprule
        \textbf{Composite Motion} & \textbf{Length} [s] & \textbf{Imitation Error} [m] \\
        \hline\hline
        Chest Open & 2.10 & $0.11 \pm 0.02$\\
        Front Jumping Jack (lower) & 1.80 & $0.16 \pm 0.03$ \\
        \hline
        Front Jumping Jack (upper) & 1.80 & $0.30 \pm 0.03$ \\
        Walk In-place & 2.10 & $0.29 \pm 0.02$\\
        \hline
        Chest Open & 2.10 & $0.10 \pm 0.01$ \\
        Squat &  1.67 & $0.09 \pm 0.01$ \\
        \hline
        Waist Twist &  3.37 & $0.15 \pm 0.04$\\
        Leg Lunge & 3.67 & $0.13 \pm 0.02$ \\
        \hline
        Hand Waving & 1.80 & $0.06 \pm 0.03$ \\
        Walk & 1.10 & $0.09 \pm 0.02$ \\
        \hline
        Punch & 1.30 & $0.11 \pm 0.02$\\
        Walk & 1.10 &  $0.10 \pm 0.01$ \\
        \hline
        Punch & 1.30 & $0.17 \pm 0.03$ \\
        Run & 0.76 & $0.14 \pm 0.01$ \\
        \hline
    \end{tabular}
    \label{tab:imit}
\end{table}

\subsection{Goal-Directed Motion Synthesis}\label{sec:goal}
\begin{figure*}[t]
    \centering
    \begin{subfigure}[b]{\linewidth}
        \includegraphics[width=\linewidth]{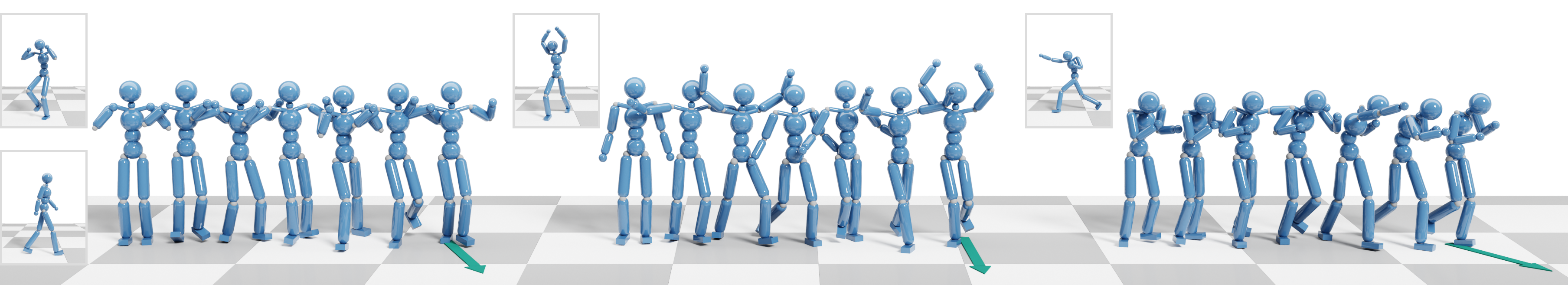}
        \caption{Tasks: Target Heading (Directional Walking with Various Upper-Body Motions)}\label{fig:heading}
    \end{subfigure}\\\vspace{0.1cm}
    
    \begin{subfigure}[b]{0.49\linewidth}
        \includegraphics[width=\linewidth]{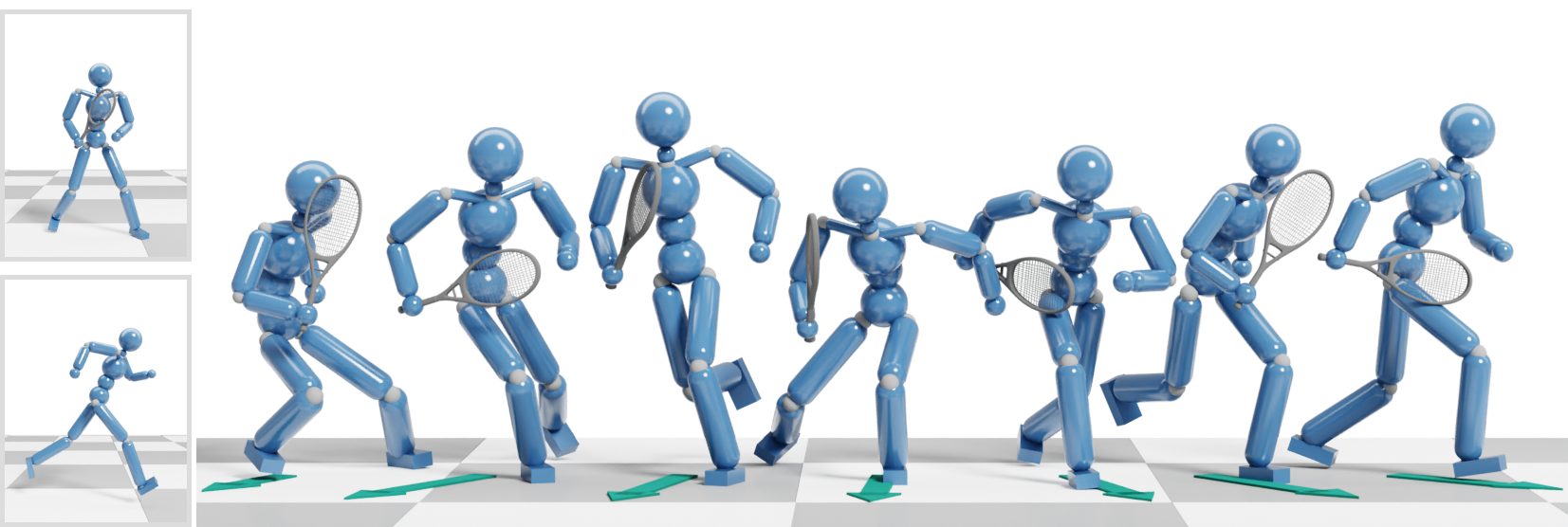}
        \caption{Task: Target Location (Run) with Tennis Racket Holding}\label{fig:target}
    \end{subfigure}\hfill
    \begin{subfigure}[b]{0.49\linewidth}
        \includegraphics[width=\linewidth]{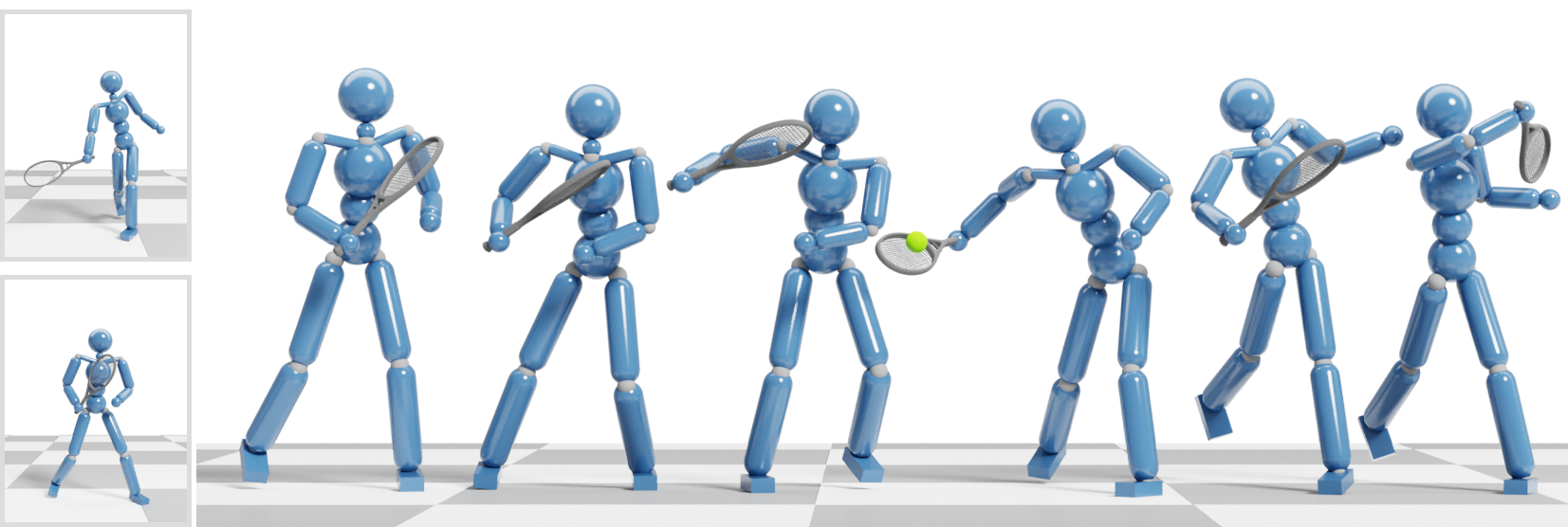}
        \caption{Task: Tennis Swing (Forehand Swing with Footwork)}\label{fig:stroke}
    \end{subfigure}

    \begin{subfigure}[b]{\linewidth}
        \includegraphics[width=\linewidth]{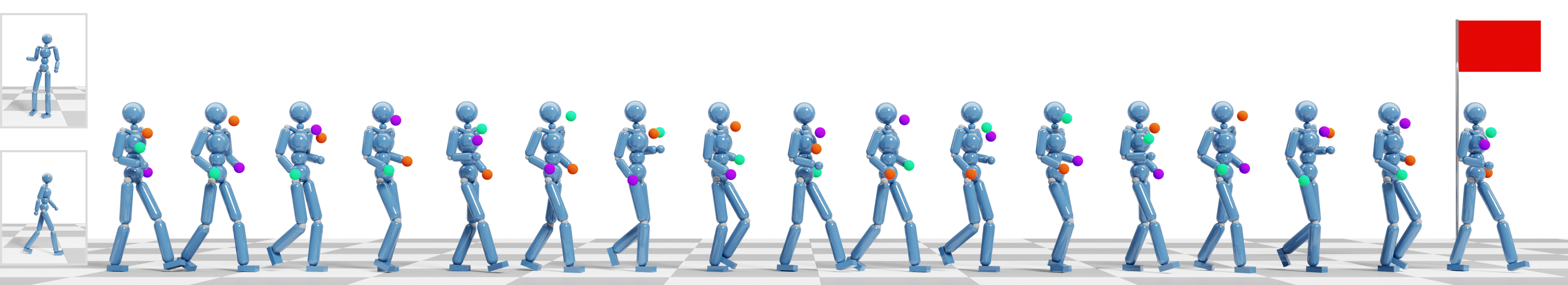}
        \caption{Task: Target Location (Walk) while Juggling }\label{fig:juggling}
    \end{subfigure}
    \caption{Motion synthesis with composite motion learning and goal-directed control. Pose snapshots shown in the small windows are captured from the reference motions.}
    \label{fig:gdc}
\end{figure*}

\begin{figure*}[t!]
    \centering
    \begin{subfigure}[b]{\linewidth}
        \includegraphics[width=\linewidth]{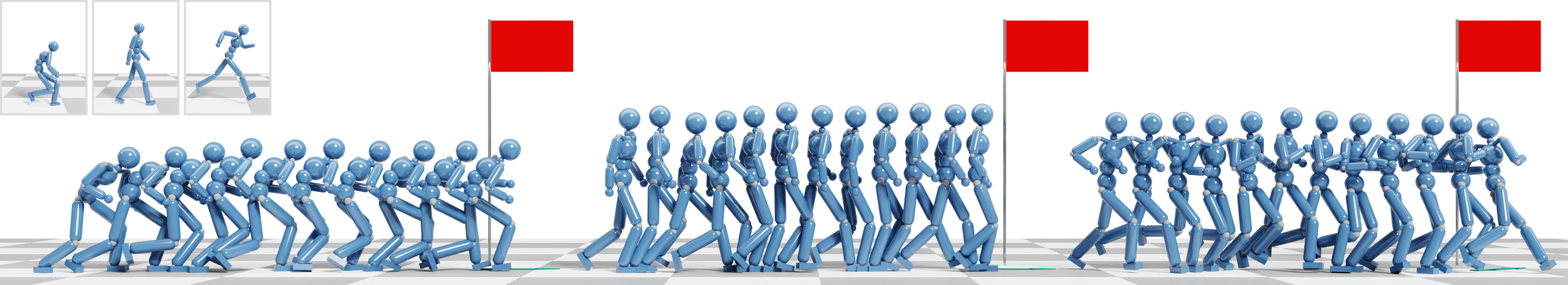}
        \caption{Meta Policy Tasks: Target Location (Crouch, Walk and Run)}
    \end{subfigure}\\\vspace{0.3cm}
    
    \begin{subfigure}[b]{\linewidth}
        \includegraphics[width=\linewidth]{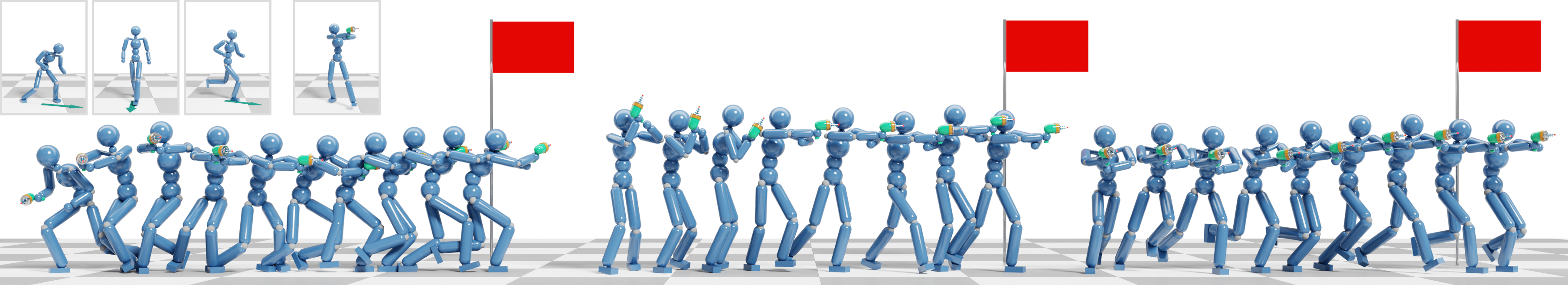}
        \caption{Incremental Learning Tasks: Directional Aiming while Location Targeting (Crouch, Walk and Run).
        }
        \label{fig:aiming}
    \end{subfigure}
    \caption{Demonstration of incremental learning tasks, where goal-directed aiming motions are added to various locomotion behaviors from the meta policies.}
    \label{fig:inc}
\end{figure*}

To test our approach with more complex tasks involving both composite motion learning and goal-directed control,
we designed five goal-directed tasks, as shown in Figs.~\ref{fig:gdc} and~\ref{fig:inc}.
In the \textit{Target Heading} and \textit{Target Location} tasks  illustrated in 
Figs.~\ref{fig:heading} and~\ref{fig:target},
the character is asked to respectively go along a target heading direction and toward a target location at a preferred speed.
Besides the goal-directed objective,
two motion imitation objectives are employed:
one is for the lower-body 
and the other one is for the upper body.
Differing from the examples shown in Fig.~\ref{fig:comp_motions} where the walking and running motions are just single, short clips containing only one gait cycle,
here we use a collection of unstructured walking and running motions as the reference for the lower body, as listed in Table~\ref{tab:imit_col}.
In the three examples shown in Fig.~\ref{fig:heading},
the upper body motions are learned from single reference motion clips,
which are chest open, jumping jack, and punch respectively,
as depicted by the small snapshots in the figure. 
In the examples shown in Fig.~\ref{fig:target},
we use the motion collection of tennis footwork as the reference for the control policy to learn how to hold the racket. 
This task is relatively harder, as the reference motions for both the upper and lower body are unstructured. 
While following the reference motions closely,
the control policies trained with our approach can effectively coordinate the character's upper and lower body poses to perform the composite motions during goal-steering navigation.

\begin{table}[t]
    \centering\small
    \caption{Motion collections used for goal-directed control.}
    \begin{tabular}{r|cc}
        \toprule
        \textbf{Motion Collection} & \textbf{\# of Clips} & \textbf{Length} [s]\\
        \hline\hline
        Crouch & 4 & 88.87 \\
        Walk & 8 & 334.07 \\
        Run &  4 & 282.87 \\
        Tennis Footwork & 2 & 31.67 \\
        Tennis Swing & 4 & 13.33  \\
        Aiming &  2 & 48.77 \\
        Juggling & 1 & 24.63 \\
        \hline
    \end{tabular}
    \label{tab:imit_col}
\end{table}

In the task of \textit{Tennis Swing},
the character is expected to hit the ball successfully with a forehand. 
The provided collection of tennis swing motions contains four trials,
where the subject performs forehand swings while standing still.
The tennis ball in our implementation is generated randomly in a small region near the character. 
As such,  
the control policy has to rely on the lower-body footwork motions to properly adjust the pose and position of the character 
relative to the tennis ball,
while it relies on the upper body swing motions to 
swing effectively and on time. 
We note that the goal-directed reward in our design only evaluates the effectiveness of hitting based on the ball's outgoing speed and destination.
The motion otherwise is decided completely by the control policy,
which leverages two discriminator ensembles to perform imitation learning for the upper and lower body respectively. 

The \emph{Tennis Swing} task is challenging, 
as it is easy for the controlled character to solely hit the ball,  but instead it is asked to do so by combining the motions from the reference collection (tennis swing for the upper body and tennis footwork for the lower body). 
The policy needs some exploration before finding a way to utilize poses from the reference motions to perform swings.
In this process,
imitation learning would fail if the policy simply tries to pursue a higher reward by simply hitting the ball. However, when
the policy is trained using our proposed multi-objective learning framework, it can balance the imitation and goal-directed objectives,
and perform forehand swings in the style of the 
reference motions.
Additionally,
while we provide only a small set of upper and lower body motions as the reference (cf. Table~\ref{tab:imit_col}),
the control policy successfully learns how to combine the motions automatically to finish the task.
In contrast,
if we just leverage full-body reference motions,
extra work is needed to generate various motions for the policy to learn. In addition, there are not
enough demonstrations for the policy to perform tennis swings correctly in a human-like style by utilizing, for example, only standing swing motions without footwork.

Figure~\ref{fig:juggling} shows another challenging composite task: \textit{Target Location} while \textit{Juggling}, where
the character needs to juggle three balls while walking to the target location.
This composite task involves four objectives: two imitation objectives and two goal-directed tasks of juggling and locomotion.
In our experiment, when a ball is relatively close to a hand, it is assumed to be caught by and attached to that hand. The ball is automatically detached from hand at a fixed interval of 20 frames. In order to perform juggling successfully and successively, after a hand releases its ball, it must catch in time a flying target ball which was thrown by the other hand. 
This task is very challenging, 
as the control policy must explore how to perform ball throwing and catching in concert with the location-targeting task. 
Besides the difficulty of throwing and catching balls, 
the juggling reference motion involves a subject balancing on a skateboard with the body swaying from side to side~\footnote{FreeMoCap Project: https://github.com/freemocap/freemocap}.
This increases the difficulty of composite motion learning to generate normal walking poses.
Differing from the other examples that use a lower and upper-body split, 
here we decouple the body parts into two groups, where one group consists of the character's arms to imitate the juggling motion and the other group includes the rest of the body parts (torso, head, pelvis, and legs) taking  the collection of walking motions as reference data. 
In such a way,
our approach can effectively eliminate the body swings
in the juggling reference motion, 
and generate composite motions with the upper body moving naturally during goal-steering navigation. 

The other goal-directed task explored in this study 
is \textit{Aiming},
in which the character holds a toy weapon in its right hand and is expected to aim it toward a specific direction.
In our experiments, that task is designed mainly to demonstrate the effectiveness of our proposed incremental learning scheme, which will be elaborated in the next section.
We refer to the Appendix 
for the details of the setup of all of our goal-directed tasks, and the supplementary video for related animation results.

\subsection{Incremental Learning}\label{sec:exp_inc_learning}

\begin{figure*}[t]
    \centering
    \hfill\includegraphics[width=0.2\linewidth]{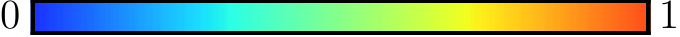}\\\vspace{0.3cm}
    
    \begin{subfigure}[b]{0.245\linewidth}
    \includegraphics[width=.45\linewidth]{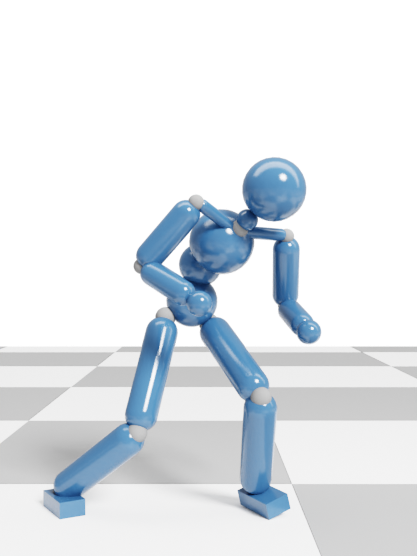}
    \includegraphics[width=.45\linewidth]{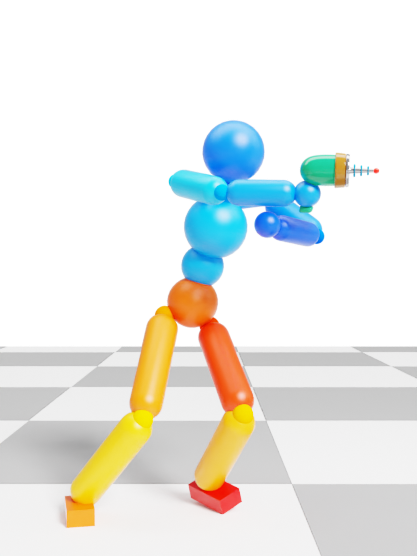}
    \caption{Aiming+Crouch}
    \end{subfigure}
    \hfill
    \begin{subfigure}[b]{0.245\linewidth}
    \includegraphics[width=0.45\linewidth]{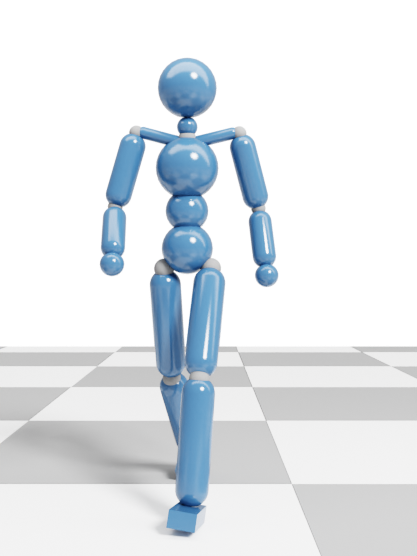}
    \includegraphics[width=0.45\linewidth]{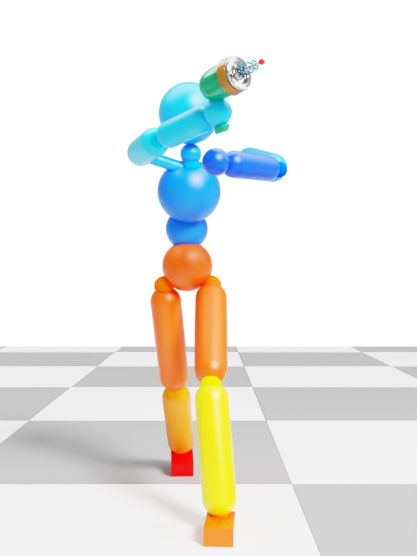}
    \caption{Aiming+Walk}
    \end{subfigure}
    \hfill
    \begin{subfigure}[b]{0.245\linewidth}
    \includegraphics[width=.45\linewidth]{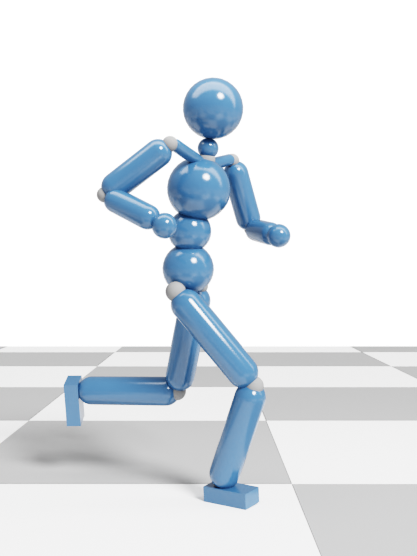}
    \includegraphics[width=.45\linewidth]{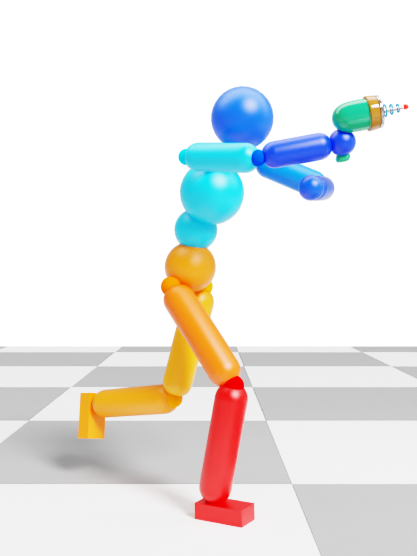}
    \caption{Aiming+Run}
    \end{subfigure}
    \hfill
    \begin{subfigure}[b]{0.245\linewidth}
    \includegraphics[width=.45\linewidth]{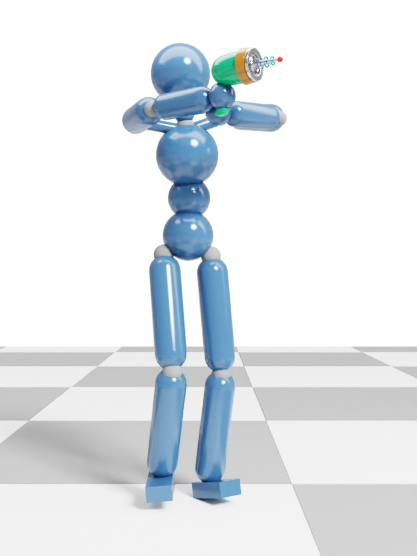}
    \includegraphics[width=.45\linewidth]{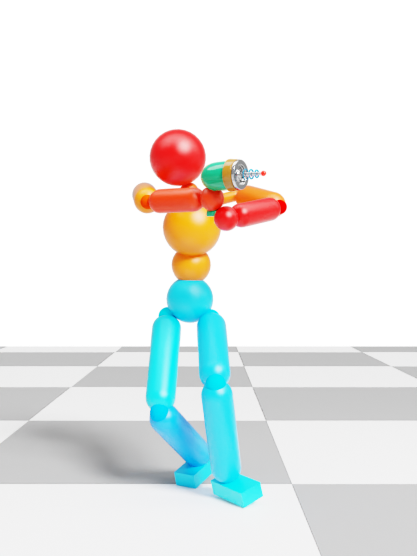}
    \caption{Crouch+AimingWalk}
    \end{subfigure}
    \caption{Visualization of the incremental learning weight $\mathbf{w}_t$ (cf. Eq.\ref{eq:pi_incremental}).
    The azure character shows the behavior from the meta policy.
    The colored character is controlled by the cooperative policy.
    The body link color identifies the weight for the associated DoF.
    The redder color represents higher weights, which means that the cooperative policy relies more on the meta policy to control the corresponding body parts of the character.
    The bluer color represents lower weights, which means that the cooperative policy mainly relies on itself to control the related body parts. 
    }
    \label{fig:inc_color}
\end{figure*}

In Fig.~\ref{fig:inc},
we show tasks used to test our proposed incremental learning scheme. 
The first row depicts three meta policies of locomotion, which are trained for the \textit{Target Location} task completely from scratch using our proposed multi-objective learning framework.
In contrast to previous examples, 
there is only one imitation objective about the full-body 
during training here, 
as shown by the snapshots on the top-left corner of the figure. 
In the 2nd row of the figure,
we show the cooperative policies that are trained by incremental learning, while reusing the pre-trained, meta policies.
In addition to the \textit{Target Location} task,
a new goal-directed task of \textit{Aiming} is introduced during training the cooperative policies.
The controlled character in this task needs to adjust its right forearm and let the toy pistol aim toward a goal direction specified dynamically.

The goal of this experiment is to demonstrate that
the cooperative policies can properly exploit the meta policies
to perform styled locomotion behaviors while quickly learning upper-body motions from the newly provided aiming reference motions, which also involve a new goal-directed task that is never seen by the meta policies.
In Fig.~\ref{fig:inc_color}, we visualize the weight vector $\mathbf{w}_t$ (cf. Eq.~\ref{eq:pi_incremental}) for each DoF by coloring the associated body link. 
The first three examples show the results obtained when we add the aiming motions to the meta policies of locomotion. 
The fourth example shows the corresponding result of adding the crouch motion to the meta policy of aiming and walking. 
As opposed to the previous meta policies, this meta policy has four objectives: two imitation objectives for the upper (aiming) and lower (walking) body  respectively, one \textit{Target Location} task and one \textit{Aiming} task. 

\begin{figure*}[t]
    \centering
    \includegraphics[width=\linewidth]{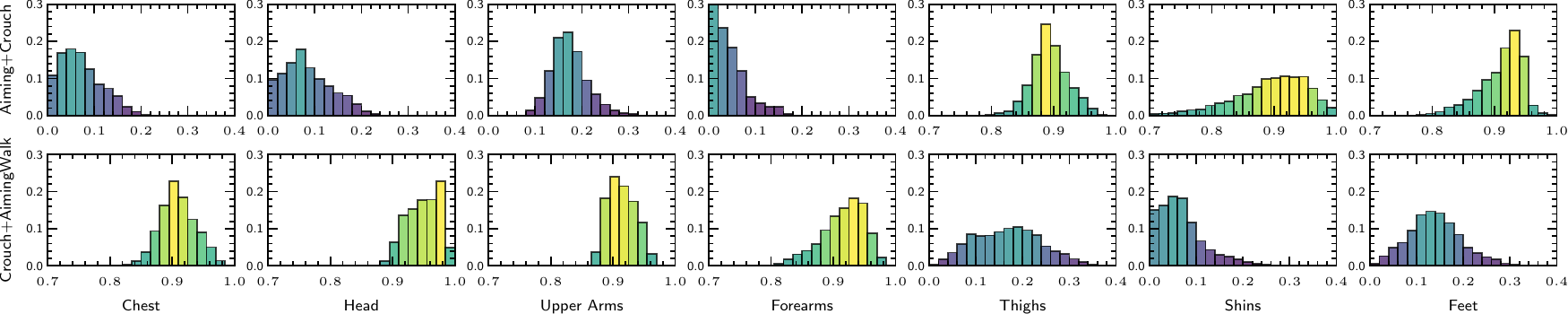}
    \caption{Distributions of the incremental learning weights $\mathbf{w}_t$ for the tasks of Aiming+Crouch and Crouch+AimingWalk (cf. Fig.~\ref{fig:inc_color}). 
    The x-axis depicts the learned weights and the y-axis shows the corresponding distribution density, normalized by the total number of samples per body part grouping. The color saturation binds the weight range for higher distribution density, with brighter colors highlighting weights greater than $0.5$.
    In the first task, the lower body is mainly controlled by the meta Crouch policy (high weights), while in the second task the AimingWalk meta policy mainly influences the upper body.}
    \label{fig:inc_weights}
\end{figure*}

As shown in the figure,
in the three Aiming+Locomotion tasks where the meta policies are pre-trained for locomotion, the cooperative policies rely more on the meta policy for lower-body actions and control the upper-body parts for aiming primarily by themselves.  
In contrast, in Crouch+AimingWalk,
we want the cooperative policy to replace the walking motions from the meta policy with crouching 
while keeping the upper-body motion of aiming. Here, as can be seen in the fourth case of the figure,
the cooperative policy exploits the meta policy to perform aiming actions but performs crouching mainly on its own. 
In Fig.~\ref{fig:inc_weights}, we also plot the distribution of weights based on the collection of 5,000 consecutive frames from the  Aiming+Crouch and Crouch+AimingWalk tasks. 
The statistical results are consistent with the above studied cases.

\begin{figure*}[t]
    \centering
        \includegraphics[width=0.99\linewidth]{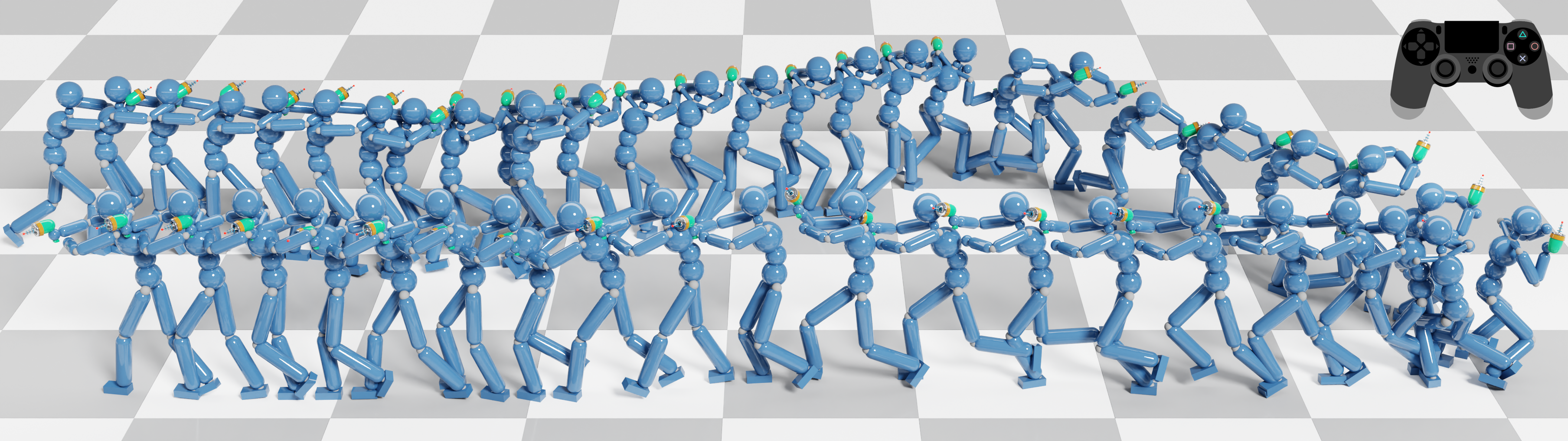}
        \caption{Interactive control of switching between walking, crouching and running for location targeting while aiming.}
    \label{fig:inter_ctrl}
\end{figure*}

As an additional experiment, in Fig.~\ref{fig:inter_ctrl},
we show that control policies trained with our approach can support the interactive control scheme proposed by~\citet{iccgan2021}.
In this experiment, we let the character 
perform a variety of locomotion styles by switching the three trained Aiming+Locomotion policies interactively
in response to external control signal provided by the user,
and navigate to and aim at the target directions specified by the user dynamically. 

\subsection{Ablation Studies}\label{sec:ablation}
\begin{figure}[t]
    \centering
    \includegraphics[width=\linewidth]{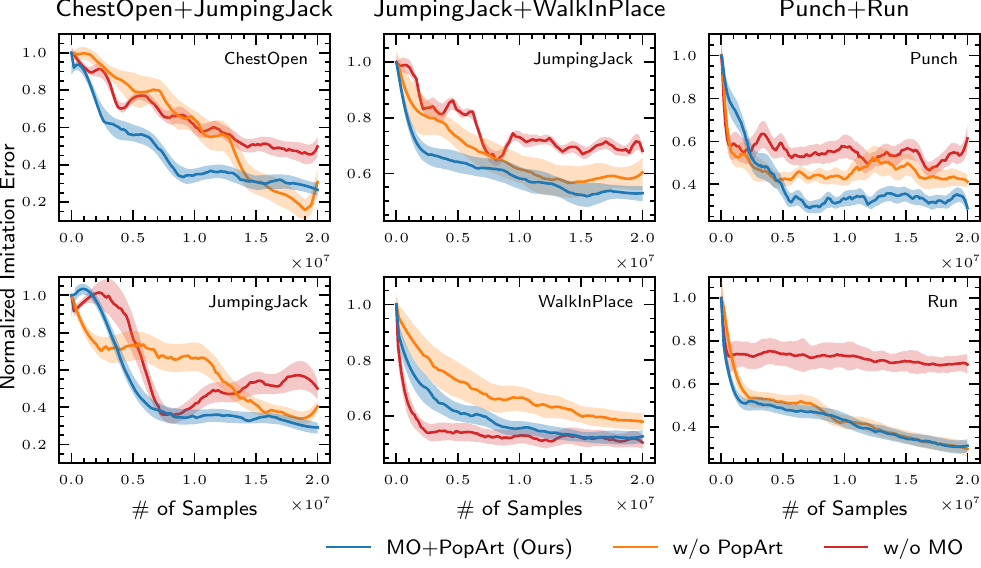}
    \caption{Learning performance on tasks of composite motion learning from two single-clip reference motions, which are illustrated in Fig.~\ref{fig:comp_motions}. "MO" stands for the proposed multi-objective learning framework detailed in Section~\ref{sec:mo_learning}. 
    Colored regions denote mean values $\pm$ a standard deviation based on 10 trials.}
    \label{fig:ablation_imitation}
\end{figure}

We refer to the previous literature of ICCGAN~\cite{iccgan2021} for ablation studies with respect to each component in the employed GAN-like structure for motion imitation, and to~\cite{peng2021amp,iccgan2021} for related analyses on the robustness of control policies trained using GAN-like structures combined with reinforcement learning. 
Here, we focus on the studies of the proposed multi-objective learning framework and incremental learning scheme.

In Fig.~\ref{fig:ablation_imitation},
we compare the performance of our proposed  multi-objective (MO) learning framework
to two baselines using three composite motion learning tasks from Section~\ref{sec:motion_decoupling}. 
The first baseline leverages our MO learning framework but does not make use of \emph{PopArt} to normalize the value targets of each critic (w/o PopArt). The second baseline simply adds the rewards from the two discriminators together and models the composite motion learning task as a typical reinforcement learning problem (w/o MO). Both baselines are trained with our motion decoupling scheme described in Section~\ref{sec:motion_decoupling} and simultaneously leverage two discriminators, one for the upper-body motion and one for the lower body. 
As can be seen from the figure, 
it is hard for "w/o MO" to balance the learning of the two reference motions.
For example, in the ChestOpen+JumpingJack task,
as the upper-body (ChestOpen) imitation error goes down, the lower-body (JumpingJack) error increases;  
in the Punch+Run task,
the policy almost gives up on learning how to run, focusing on punching without too much success. 
In contrast, when leveraging  our MO framework either with or without \textit{PopArt},  
the imitation errors of the upper and lower body show similar and stable trends,  keep decreasing as the training goes on. 
Additionally, the introduction of \textit{PopArt} typically facilitates better training, 
allowing for faster convergence speed, lower imitation error, and more robust training achieving similar performance across different trials.

\begin{figure}[t]
    \centering
    \includegraphics[width=\linewidth]{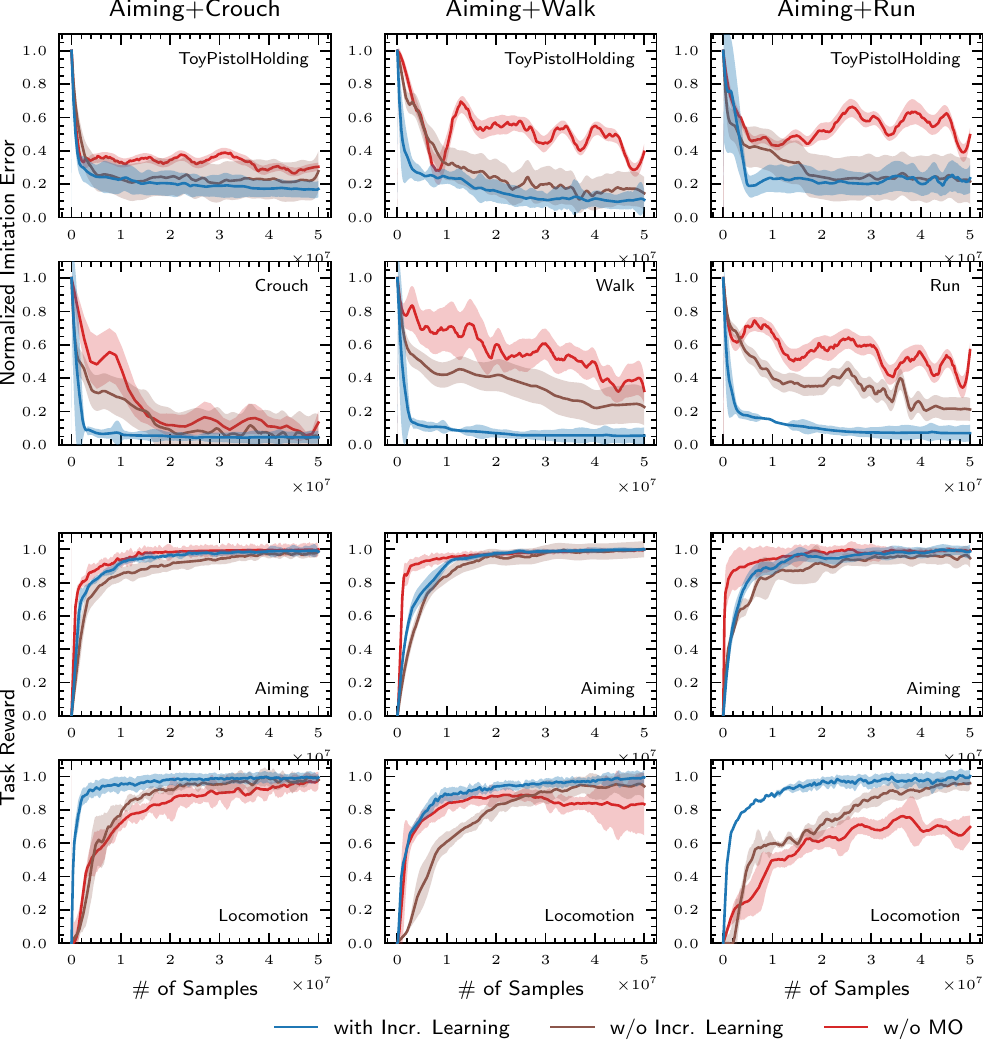}
    \caption{
    Learning performance on three composite tasks where each task combines learning from two partial motions while accomplishing two goal objectives. Multi-objective learning in an incremental manner leads to sample-efficient training allowing for high-fidelity composite motion synthesis with goal-directed control. Colored regions denote mean values $\pm$ one standard deviation based on 10 trials.}
   \label{fig:ablation_inc_learning}
\end{figure}

Figure~\ref{fig:ablation_inc_learning} shows the performance of 
our MO approach with and without exploiting the proposed incremental learning scheme. We also provide comparisons with the "w/o MO" baseline. 
The tested tasks have four objectives, as described in Section~\ref{sec:exp_inc_learning}: two imitation objectives for the upper and lower body respectively, one \textit{Target Location} task for the locomotion and one \textit{Aiming} task.
In the cases 
using incremental learning, we employed a pre-trained, locomotion policy as the meta one. 
Consistent with the previous ablation study, we can see that the "w/o MO" baseline struggles to balance the different objective terms. Here, 
the character quickly achieves a high reward for the goal-directed \textit{Aiming} task (3rd row) 
but fails to 
complete other objectives, and in particular to 
account for the motion style provided by the imitation reward terms. For example, the controlled character holds the toy pistol in an unnatural way compared to the demonstrations in the provided reference motions as indicated by the high imitation error (1st row). 
While such issues are successfully resolved by our proposed MO framework, learning in a non-incremental way leads to sample inefficient training as compared to learning by leveraging a meta policy. 
Besides slow speed of convergence, non-incremental training can be time consuming for challenging multi-objective tasks. 
For example, in the Aiming+Run task,
while the case with incremental learning only needs 1.5
hours to finish the training by using about 20 million samples,
the non-incremental cases need about 20 more hours for training and will consume 
about 300 million more samples to achieve a similar performance.

\section{Limitations and Future Work}
\label{sec:limitations}
We present a technique for training composite-motion controllers using a multi-objective learning framework that is capable of combining multiple reference examples and task goals to control a physically-simulated character. 
We demonstrate that our approach can generalize to a large number of examples based on the availability of reference data.  Likewise, we show its ability to accomplish simultaneous goal-driven tasks such as aiming at specific targets and moving to a target location with different locomotion styles. Furthermore, we can interactively control such character's actions, 
pushing the boundary of what is capable for physics-based characters to date.

Of course, there is still more to explore in this space. 
Our system is currently not well-equipped to handle behaviors which include multiple phases, as
the imitation is not phase-locked in any fashion and our discriminators do not distinguish between different stages of an activity. Exploring the potential to add a state machine with state transitions could aid in this capacity~\cite{takufsm}. Another shortcoming of the approach presently is that we do not account for variation across the humans that recorded the motion clips.  This implies that we are introducing bias in the imitation process that may degrade the final quality of the animation.   
As is, the system is able to make adjustments automatically as needed based on the physical characteristics of the behavior but it cannot distinguish errors that are more stylistic.

In its current form, our system can not create new composite activities without performing additional training.  A possible direction for future work is aimed at sidestepping this limitation to directly combine preexisting policies and greatly improve the scalability of trained controllers.  That is, to train two (or more) policies independently and combine them at runtime to create a composite motion. 
Finally, 
in human motion, composite behaviors go beyond an anticipated split, e.g. the lower and upper body, which is one of the modest underlying assumptions in our current implementation.  Instead, humans may enlist body parts and release them fluidly.
For example, a well-trained martial artist changes the use of appendages quickly in fighting sequences.  We wish to explore this direction in future investigations and believe that our proposed multi-objective learning framework can provide the foundation for such future endeavors. 

\begin{figure}
    \centering
    \includegraphics[width=\linewidth]{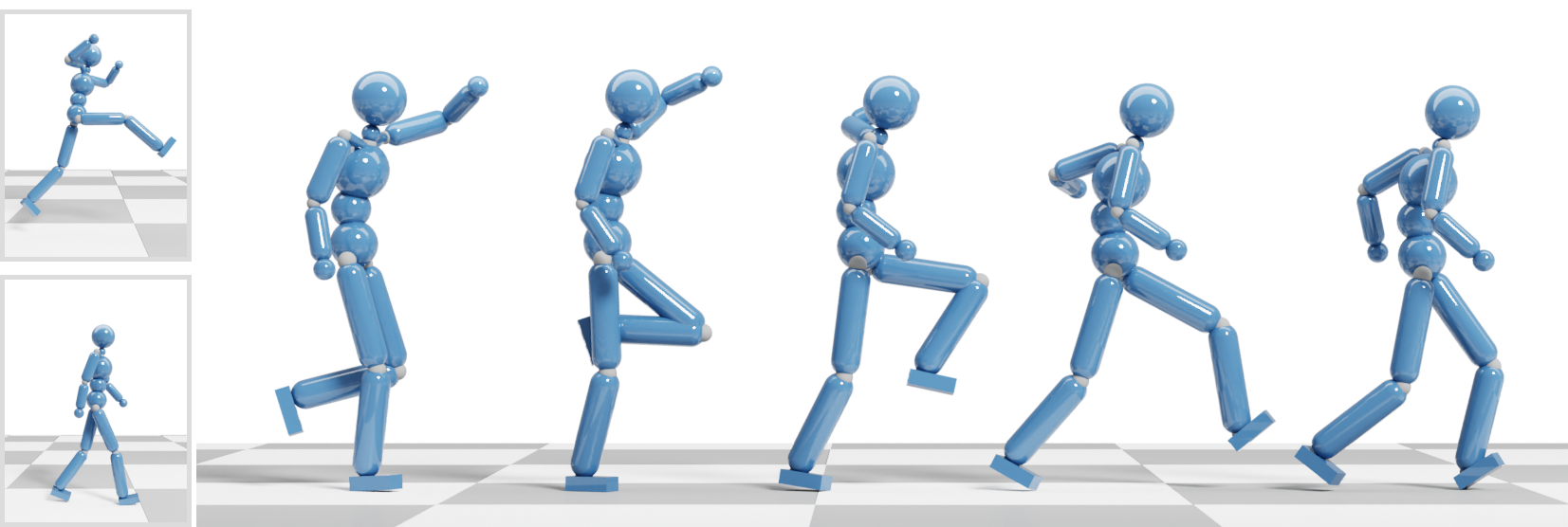}
    
    \includegraphics[width=\linewidth]{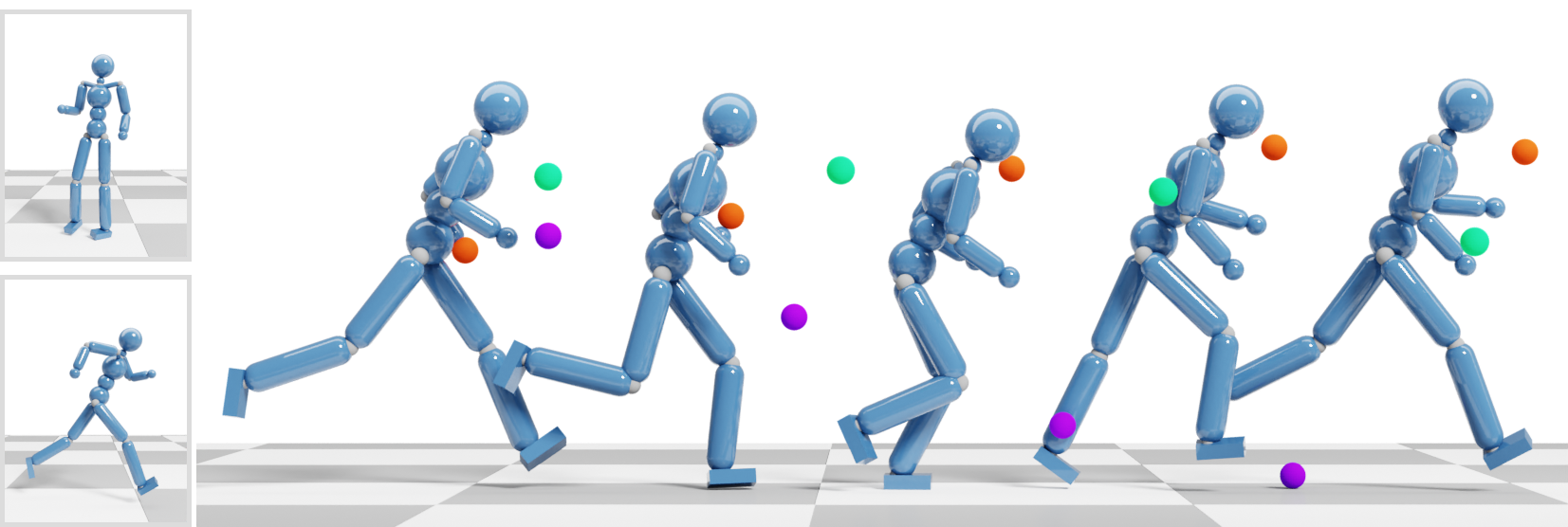}
    \caption{Failure case study. 
    Top: The character's body is bisected into a left and right group, imitating walking and jumping respectively.
    Bottom: Juggle while running. 
    }
    \label{fig:failure}
\end{figure}

Although we employed an upper and lower body split in most of our experiments, there is nothing tied to this body decoupling scheme 
except that it is a practical general choice for deploying the limbs of the whole body.
Currently, as long as the subtasks are compatible, our system is capable of combining motions along other body splits. 
For instance, in the
\textit{Juggling+TargetLocation}
example discussed in Section~\ref{sec:goal}, the trained policy controls the arms for juggling and the rest of the body for walking. 
Our approach may fail if, for example, the lower limbs are separated due to the requirements of physical balance. 
As an example, in Fig.~\ref{fig:failure}, we show a failure case where the body is bisected into a left/right split and asked to imitate walking and jumping motions respectively.
Such a composite motion is not well-defined, even for humans.
We can see that though not falling down, the simulated character cannot imitate the two motions accurately, and instead performs an in-between motion where the character neither jumps up nor walks in an expected fashion.

In Fig.~\ref{fig:failure}, we also show another failure case where running reference motions with an average speed of around $3.5m/s$ are provided for the \textit{Juggling+TargetLocotion} task.
With the difficulty of 
juggling while moving at this higher speed,
this example is significantly more challenging than the one shown in Fig.~\ref{fig:juggling}.
Even though we are able to synthesize the composite motions,
the simulated character cannot juggle the balls successfully under these
conditions. 
Currently, our approach cannot identify if a composite motion is compatible on its own, and instead, it relies on a human to combine behaviors with some domain knowledge about the affinity 
of the mixing and the feasibility of associated goal-directed tasks. Automating this would be a great direction for future work.

\begin{acks}
This work was supported by the National Science Foundation under Grant No. IIS-2047632 and by Roblox.
We would like to thank Rokoko~\footnote{https://www.rokoko.com} 
for providing mocap data for this project.
\end{acks}

\bibliographystyle{ACM-Reference-Format}
\bibliography{main.bib} 

\setcounter{page}{1}

\appendix

\section{State and Action Representation}\label{sec:state}

Given the definition in Eq.~\ref{eq:s},
we have the character state vector
$\mathbf{s}_t \in \mathbb{R}^{(n+1) \times N_\text{link} \times 13}$,
which includes all body links' positions, orientations, and linear and angular velocities of the simulated character in the last $n+1$ frame from $t-n$ to $t$.
To ignore the global coordinate,
we assume that the ground height is $0$, and
all the body links' states are localized based on the position and heading direction of the character's root link (pelvis) at the last frame $t$.
Similarly, if a goal state $\mathbf{g}_t$ is provided,
we localize the position and direction state in $\mathbf{g}_t$ using the same coordinate system with $\mathbf{s}_t$.
During our experiments, if multiple goal-directed tasks are involved,
we simply concatenate goal states from all the tasks together as the representation of $\mathbf{g}_t$.
We refer to the Appendix in the supplementary material for the representation of $\mathbf{g}_t$ in our designed goal-directed tasks.

The action $\mathbf{a}_t$ is a set of target postures fed into the PD servo. 
Therefore, we have $\mathbf{a}_t \in \mathbb{R}^{N_\text{dof}}$
where $N_\text{dof}$ is the total degrees of freedom (DoF) in the character model.
$\mathbf{a}_t$ is assumed to be normalized by the valid movement range of each DoF
but without upper and lower bounds applied. 
The observation space $\mathbf{o}_t^i$ for discriminators is similar to $\mathbf{s}_t$.
However, 
we keep only
body links' positions and orientations,
and the discriminators rely on the pose trajectory of $\mathbf{o}_t$ to ensure that the visual velocities between two frames are consistent with the reference motions.
As such, we have 
$\mathbf{o}_t^i \in \mathbb{R}^{(n_i+2) \times N_\text{link}^i \times 7}$
where $n_i+2$ is the number of observed frames as defined in Eq.~\ref{eq:ob}.
$\mathbf{o}_t^i$ is localized depending on its characteristics.
For lower body parts, their motions often involve the character's spatial movement.
Therefore, we follow the definition of $\mathbf{s}_t$,
and use a local coordinate defined by the root pose at the last observed frame.
For upper-body motions, however,
we typically care more about the body parts' local poses related to a specific parent body link.
Therefore,
we use a framewisely defined local system based on the parent link's pose 
such that the global-space displacement and rotation controlled by the lower body are ignored.
In our implementation,
for upper-body motions, we choose pelvis as the parent link;
and for arm only motions, we choose torso as the parent.

The observation sampled from the reference motions, i.e. $\mathbf{\tilde{o}}_t^i$, is defined the same as $\mathbf{o}_t^i$.
However, instead of performing sampling at a fixed frame rate identical to the control policy's working frequency (30Hz in Fig.~\ref{fig:overview}),
we do sampling with dynamic interval $\Delta t = \beta T$ where $T = 1/30s$ is the time interval between two frames during simulation and $\beta \sim \textsc{Uniform}(0.8, 1.2)$.
In such a way, we scale the reference motion temporally within a small range,
for better combining motions from multiple reference sources with inconsistent pace.
To keep the motion stable,
$\Delta t$ differs among multiple times of sampling but is identical for the $n_i+2$ frames of one sample.

\section{Task Environment Setup}\label{app:task}
\subsection{Task: Target Heading}
The goal-directed reward is defined as
\begin{equation}
    r_t = \langle \dot{\mathbf{x}}_{t+1}^\text{root} / \vert\vert \dot{\mathbf{x}}_t \vert\vert, \mathbf{g}_t \rangle, 
\end{equation}
where $\dot{\mathbf{x}}_{t+1}^\text{root}$ is the horizontal displacement of the character's root link from the frame $t$ to $t+1$.
The goal state $\mathbf{g}_t \in \mathbb{R}^2$ is a unit vector representing the target heading direction, which is randomly sampled every 30 frames ($1s$).

\subsection{Task: Target Location}
The goal-directed reward is defined as
\begin{equation}
    r_t = \begin{cases}
        \exp(-3\vert\vert\dot{\mathbf{x}}_{t+1}^\text{root}/T - \mathbf{v}_t^\ast\vert\vert^2/\vert\vert \mathbf{v}_t^\ast \vert\vert^2) & \text{if } \vert\vert\mathbf{x}_{t+1} - \mathbf{p}_\text{goal}\vert\vert > R\\
        1 & \text{otherwise},
    \end{cases}
\end{equation}
where $R=0.5$ is the goal radius of the target location, $T = 1/30s$ is the time interval between two frames,
$\dot{\mathbf{x}}_{t+1}^\text{root}/T$ denotes the horizontal velocity of the character's root link from the frame $t$ to $t+1$, 
and $\mathbf{v}_t^\ast$ is the target velocity with a preferred speed and a direction toward the target goal location.

The goal state $\mathbf{g}_t \in \mathbb{R}^4$ includes a 2D unit vector representing the direction to the target location, the horizontal distance from the character to the goal, i.e. $\vert\vert\mathbf{x}_t^\text{root} - \mathbf{p}_\text{goal}\vert\vert$, and the preferred speed $\vert\vert \mathbf{v}_t^\ast \vert\vert$.
The preferred speed is sampled from $[1, 1.5]$ in the unit of $m/s$ for crouching and walking motions, and from $[1, 3]$ for running.
The goal direction is sampled from $[0, 2\pi)$.
A timer variable is sampled from $[3, 5]$ in the unit of $s$ for crouching and walking motions, and from $[2, 3]$ for running.
We use these three goal variables to obtain the target location.
As such, we can perform speed control during the location targeting.

\subsection{Task: Aiming}
The goal-directed reward is defined as
\begin{equation}
    r_t = \begin{cases}
        \exp(-2\vert\vert\mathbf{d}_t^\text{forearm} - \mathbf{g}_t\vert\vert^2) & \text{if aiming is activated}\\
        \textsc{Clip}(\langle \mathbf{d}_t^\text{forearm}, \mathbf{u}^\text{ref}\rangle, 0, 0.8)/0.8 & \text{otherwise}
    \end{cases}
\end{equation}
where $\mathbf{d}_t^\text{forearm} \in \mathbb{R}^3$ is a unit vector representing the direction of the right forearm from the elbow to the hand, and $\mathbf{u}^\text{ref}$ is a unit vector representing the up axis of the world space.
In our implementation, the toy pistol is fixed on the right hand, which is linked to the right forearm with a fixed joint.
Therefore, we use the direction of the right forearm as the aiming direction.
When the aiming action is not activated,
we use the 2nd reward term to encourage the character to lift its arm and hold the gun up without aiming anything.

The goal state $\mathbf{g}_t \in \mathbb{R}^3$ is a unit vector representing the target aiming direction.
We let $\mathbf{g}_t = \mathbf{0}$ if the aiming action is not activated.
When combined with the target location task,
aiming is deactivated if the character is close to the goal, i.e. $\vert\vert\mathbf{x}_t - \mathbf{p}_\text{goal}\vert\vert \leq R$.
$\mathbf{g}_t$ is sampled with an elevation angle in range of $[0, \pi/6]$ and azimuth angle in $[0, \pi/4]$.

\subsection{Task: Tennis Swing}
The goal-directed reward is defined as
\begin{equation}
    r_t = \begin{cases}
       1.2 + \vert\vert \mathbf{v}_\text{out} \vert\vert / 10 & \text{if ball was hit and $d_\text{fall} = 0$}\\
       r_t^\text{pose} + 0.5 \exp(-0.1 d_\text{fall}^2) & \text{if ball was hit but  $d_\text{fall} > 0$}\\
       r_t^\text{pose} & \text{otherwise}\\
    \end{cases} 
\end{equation}
where
\begin{equation}\begin{split}
    & r_t^\text{pose} = 0.2 r_t^\text{shoulder} + 0.5 r_t^\text{racket}, \\
    & r_t^\text{shoulder} = \exp(-\max(\vert\vert \mathbf{p}_t^\text{shoulder} - \mathbf{p}_t^\text{ball}\vert\vert-1, 0)^2), \\
    & r_t^\text{racket} = \exp(-5\vert\vert \mathbf{p}_t^\text{racket} - \mathbf{p}_t^\text{ball} \vert\vert^2). \\
\end{split}
\end{equation}
$\mathbf{p}_t^\text{shoulder}$ is position of the character's right shoulder and $\mathbf{p}_t^\text{racket}$ is the position of the racket.
To emulate the tennis court,
we consider a valid ball falling region with dimension $12m \times 11m$, which is $6m$ ahead of the initial position of the tennis ball along the x-axis.
$d_\text{fall}$ is the distance from the ball's falling point to this region.
We let $d_\text{fall} = 0$ if the ball will fall or fell in the target region. 
$d_\text{fall}$ is estimated by a simple projectile model based on the linear velocity of the ball without considering any friction or air resistance,
but updated at every simulation step in order to get an accurate estimation.
$\vert\vert \mathbf{v}_\text{out} \vert\vert$ is the outgoing speed of the tennis ball when it was hit.
The purpose of using $r_t^\text{shoulder}$ is to encourage the character to approach the tennis ball but not necessarily when the distance is less than $1m$
such that the character can have enough space to swing the racket,
rather than keeping moving close to the ball.

The goal state $\mathbf{g}_t \in \mathbb{R}^4$ includes a 3D vector representing the position of the ball $\mathbf{p}_t^\text{ball}$, 
and a scalar identifying the heading direction of the character's root link.
The heading direction in $\mathbf{g}_t$ is used to identify the direction of x-axis toward which the ball is expected to be hit.
We let $\mathbf{p}_t^\text{ball} = \mathbf{0}$ when constructing $\mathbf{g}_t$ if the ball was hit.

\subsection{Juggling}
The goal-directed reward is defined as
\begin{equation}
    r_t = 0.5r_t^{\text{hand}, \text{left}} + 0.5r_t^{\text{hand}, \text{right}}
\end{equation}
where $r_t^{\text{hand}, \text{left}}$ and $r_t^{\text{hand}, \text{right}}$ are defined identically but evaluate the performance of the left hand and right hand respectively.
For each hand-related reward, we define
\begin{equation}
    r_t^{\text{hand}} = \begin{cases}
       r_t^{\text{throw}} & \text{if $t\bmod\tau = 0$} \\
       0.1r_t^{\text{height}} + 0.9r_t^{\text{distance}} & \text{otherwise}\\
    \end{cases} 
\end{equation}
where $\tau$ is the time interval between two trials of ball throwing and
\begin{equation}\begin{split}
    & r_t^{\text{throw}} = \exp(-5(v_t^{\text{ball}}/V^{\text{ball}} - 1)^2), \\
    & r_t^{\text{height}} =\exp(-20(h^{\text{ball}} - h_t^{\text{hand}})^2), \\
    & r_t^{\text{distance}} = 0.9\exp(-20d_t^2) + 0.2\exp(-d_t^2).
\end{split}\end{equation}

As stated in Section~\ref{sec:goal}, 
we employ an automatic catch-and-throw mechanism
where a ball is considered caught by a hand and is fixed to that hand if it is close enough, and will be detached (thrown) automatically at a fixed time interval $\tau$ 
between two trials of throwing.
The target ball for a hand is decided using a \textit{cascade} juggling pattern. 
In the reward function, $r_t^\text{throw}$ measures the performance of ball throwing and is computed only at the frame where a ball is thrown.
$v_t^{\text{ball}}$ is the vertical velocity of the thrown ball and $V^{\text{ball}}$ is the preferred vertical thrown velocity.
The preferred velocity is obtained by assuming that the thrown ball will be caught at the same height where it is thrown and at a dwell time $t_d$ before the next time the catching hand performs a thrown. 
In our experiment, we set $\tau = 2/3s$ (20 frames) with a preferred dwell time $t_\text{d} = 0.4s$ (12 frames) and set the number of balls $N_\text{ball} = 3$. 
Given the gravity $g=9.81m/s^2$, this leads to a preferred velocity
\begin{equation}
    V^{\text{ball}} = 0.5g(\frac{\tau}{2}N_\text{ball} - t_\text{d}) = 2.94m/s.
\end{equation}
The height-related reward term $r_t^{\text{height}}$ measures the error between the hand's vertical position ($h_t^{\text{hand}}$) and the target ball's height when it was thrown ($h^{\text{ball}}$).
It encourages the control policy to throw and catch a ball at the same height.
We let $r_t^{\text{height}}=1$ if the target ball was caught by the hand already.
The distance-related reward $r_t^{\text{distance}}$ measures the distance error between the hand and the target ball.
We estimate the ball's vertical movement trajectory using a simple projectile model taking into account only the ball's vertical linear velocity and gravity.
The distance $d_t$ is defined as the distance between the hand and the target ball if the hand is above the estimated trajectory, i.e. when the hand is unable to catch the ball at the current hand height, or just the horizontal distance otherwise.
As such, $r_t^{\text{distance}}$ ignores the vertical ball-hand distance if the hand is able to catch the ball at its current height,
and thus prevents the hand from aggressively moving toward the ball vertically. 

The goal state $\mathbf{g}_t \in \mathbb{R}^{19}$ includes the three balls' states (position and linear velocity) and a timer variable counting the time left before the next throwing of the ball by one hand. The ball states are in the order of the left-hand target ball, the right-hand target ball, and the other ball.
For a caught target ball, we let the corresponding state be zero.

\begin{table}[t]
\centering
\caption{Hyperparameters}
\begin{tabular}{lc}
    \toprule
    \textbf{Parameter} & \textbf{Value}\\
    \midrule
    policy network learning rate & $5 \times 10^{-6}$\\
    critic network learning rate & $1 \times 10^{-4}$\\
    discriminator learning rate & $1 \times 10^{-5}$\\
    reward discount factor ($\gamma$) & $0.95$ \\
    GAE discount factor ($\lambda$) & $0.95$ \\
    surrogate clip range ($\epsilon$) & $0.2$ \\
    gradient penalty coefficient ($\lambda^{GP}$) & $10$ \\
    number of PPO workers (simulation instances) & $512$ \\
    PPO replay buffer size & $4096$ \\
    PPO batch size & $256$ \\
    PPO optimization epochs & $5$ \\
    discriminator replay buffer size & $8192$ \\
    discriminator batch size & $512$ \\
  \bottomrule
\end{tabular}
\label{tab:hyper}
\end{table}

\section{Hyperparameters}\label{app:hyper}
The hyperparameters used for policy training is listed in Table~\ref{tab:hyper}.
Half of the samples for discriminator training are from the simulated character and half are sampled from the reference motions.
The character state horizon $n+1$ is chosen as 4, 
and the discriminator observation horizon $n_i+2$ is 3 for aiming motions and 5 for other motions.
The objective weight $\omega_k$ in Eq.~\ref{eq:policy_loss_one_pass} is 0.5 shared equally by all goal-related objectives.
In the Juggling with Target Location task, given the difficulty of ball catching, the juggling task is assigned a weight of 0.6, the locomotion task has a weight of 0.1, and the imitation tasks account for the remaining weight with a ratio of $1:4$ for juggling and walking motion imitation.
In the Aiming+Locomotion task, the upper-body motion of aiming has a weight of 0.2 and the lower-body motion has a weight of 0.3.
On the other tests, besides the weights taken by the goal-related objectives,
the remaining weight is shared equally by the imitation objectives.

\end{document}